\documentclass[aps,prd,twocolumn,nofootinbib,superscriptaddress,10pt]{revtex4-2}

\usepackage[utf8]{inputenc}
\usepackage[T1]{fontenc}

\usepackage{graphicx}
\usepackage{amsmath,amssymb,amsfonts,mathtools}
\usepackage{amsthm}
\usepackage{mathrsfs}
\usepackage{bm}
\usepackage{bbm}
\usepackage{physics}
\usepackage{slashed}
\usepackage{cancel}

\usepackage{tabularx}
\usepackage{array}
\usepackage{multirow}
\usepackage{booktabs}
\usepackage[section]{placeins}
\usepackage{algorithm}
\usepackage{algorithmicx}
\usepackage{algpseudocode}
\usepackage{listings}

\usepackage[title]{appendix}
\usepackage{textcomp}
\usepackage{comment}
\usepackage{xspace}
\usepackage{xcolor}

\usepackage{hyperref}
\hypersetup{
    colorlinks=true,
    linkcolor={red!50!black},
    citecolor={blue!50!black},
    urlcolor={blue!80!black}
}

\numberwithin{equation}{section}
\newcommand{\BSMArt}{\textsc{BSMArt}\xspace}
\newcommand{\SARAH}{\textsc{SARAH}\xspace}
\newcommand{\SPheno}{\textsc{SPheno}\xspace}

\newcommand{\HiggsTools}{\textsc{HiggsTools}\xspace}
\newcommand{\MicrOMEGAs}{\textsc{MicrOMEGAs}\xspace}
\newcommand{\SModelS}{\textsc{SModelS}\xspace}

\newcommand{\beq}{\begin{equation}}
\newcommand{\eeq}{\end{equation}}

\newcommand{\UB}{U(1)_{\tilde B}}

\newcommand{\tB}{\tilde B}

\begin{document}

\title{An anomaly-free baryonic completion of the Standard Model}

\author{Karim Benakli}
\email{kbenakli@lpthe.jussieu.fr}

\author{Mark Goodsell}
\email{goodsell@lpthe.jussieu.fr}

\author{Arno Goudeau}
\email{goudeau@lpthe.jussieu.fr}

\affiliation{Sorbonne Universit\'e, CNRS, Laboratoire de Physique Th\'eorique et Hautes Energies (LPTHE), F-75005 Paris, France}

\date{\today}

\begin{abstract}
We construct an explicit anomaly-free completion of a gauged baryon-number
symmetry. The new gauge factor is \(\UB\): when restricted to the
Standard Model fields its charge \(\tB\) coincides with ordinary baryon
number, whereas in the full theory it becomes an extended baryonic symmetry
acting on both the Standard Model and a secluded sector. We show that anomaly
cancellation alone leaves a broad class of solutions, but that requiring
consistent symmetry breaking, viable decays of the exotic states, the absence
of stable coloured relics, a neutral dark matter candidate protected by a
\(Z_2^{\rm DM}\) parity, and improved high-scale gauge-coupling unification
strongly constrains the structure of the model. These requirements necessitate
enlarging the minimal secluded fermion sector by additional scalar fields and
Yukawa interactions, giving rise to a light pseudoscalar that plays an
important role in the dark matter phenomenology. We implement the resulting
renormalisable model in \SARAH\ and analyse its spectrum and phenomenology
with \SPheno, \MicrOMEGAs, \SModelS, and \HiggsTools. We identify viable
regions of parameter space containing a heavy baryonic \(Z'\), an even
heavier secluded sector, a neutral dark matter candidate compatible with relic-density and direct-detection constraints, and improved
gauge-coupling evolution relative to the Standard Model. In these regions,
the relic abundance is determined either by co-annihilation with
nearby secluded states or by annihilation into light pseudoscalars from the
extended scalar sector.
\end{abstract}

\maketitle

\tableofcontents
\setcounter{footnote}{0}

\section{Introduction}
\label{sec:introduction}

Baryon number is one of the more striking features of the Standard
Model (SM). It is not imposed as a fundamental principle but emerges
as an accidental symmetry of the renormalisable Lagrangian, respected
by every renormalisable perturbative interaction. Its persistence
across such disparate phenomena as nuclear stability and the absence
of proton decay invites a natural question: might baryon number be
the low-energy remnant of an underlying local gauge symmetry,
spontaneously broken at a scale above the electroweak one yet low
enough to be within reach of foreseeable experiments? This is the
possibility we construct and study here.

Gauging the ordinary baryon number \(U(1)_B\) with SM matter alone
does not yield a consistent theory: the mixed gauge anomalies do not
cancel, and the associated gauge boson cannot be introduced without
additional chiral fermions to restore consistency. Early constructions
along these lines were given in
Refs.~\cite{Rajpoot:1989yq,Carone:1995pu}, while more recent work
connecting local baryon number, anomaly cancellation and dark matter
includes Ref.~\cite{FileviezPerez:2010gw} and references therein.
Extra Abelian gauge factors, kinetic mixing and hidden-sector vector
bosons have a longer history still
\cite{Okun:1982xi,Galison:1983pa,Holdom:1985ag,Fayet:1990yq,
delAguila:1995rb,Feldman:2007wj,Dobrescu:2004tq}, and the subject
continues to attract attention, as reflected in recent reviews and
experimental summaries
\cite{Fabbrichesi:2020wbt,Graham:2021gfy,Caputo:2021eaa}.

Our construction follows Ref.~\cite{Anastasopoulos:2024bxx} in its
choice of anomaly-cancelling heavy fermions. These are vector-like
under the SM gauge group but chiral under the extra Abelian symmetry,
with charges chosen such that all gauge anomalies cancel. We go beyond
that work in two respects, by imposing further phenomenological
requirements, namely improved gauge-coupling unification and a viable
dark matter candidate, and by implementing and analysing the full
renormalisable model numerically.

We denote the new Abelian gauge factor by
\begin{equation}
    \UB .
\end{equation}
On the SM fields \(\tB\) coincides with ordinary baryon number,
\[
    \tB\big|_{\rm SM}=B ,
\]
in the normalisation adopted for the benchmark model. Beyond the SM
sector, however, \(\tB\) is better understood as an extended baryonic
charge: its values on the secluded fields are fixed by anomaly
cancellation and by the renormalisable interactions needed to
generate masses and allow the heavy states to decay. The anomaly-free
completion in fact contains \(SU(3)_c\)-neutral states carrying
non-zero \(\tB\) charge, so the new symmetry cannot simply be read as
ordinary baryon number extended to every field in the theory. The
associated massive gauge boson may reasonably be called a baryonic
\(Z^\prime\), though its couplings are dictated throughout by the
requirements of the full anomaly-free completion.

The anomaly equations admit many solutions, and anomaly cancellation
alone does not fix the secluded sector uniquely; phenomenological
considerations must do the rest of the work. We require a stable dark
matter candidate, protected by a discrete parity, consistent with the
observed relic abundance (see
Ref.~\cite{Cirelli:2024ssz} for a recent review), and we favour matter
content that improves the evolution of the SM gauge couplings towards
a common value at high energy. We use the word
\emph{unification} in this looser sense: not a Grand Unified Theory
with a simple gauge group fixing all charge normalisations, but the
weaker requirement that the gauge couplings converge towards a common
scale, of the kind that can arise in string-inspired ultraviolet
completions where several gauge factors are governed by a common
modulus. The model studied here is non-supersymmetric; a
supersymmetric realisation of the same construction is under
investigation and will be presented elsewhere.

Choosing a secluded matter content that satisfies these requirements
is not a simple matter. A full mirror of the SM families, vector-like
under the SM gauge group but chiral under \(\UB\), suggests itself as
a natural first attempt, but in a non-supersymmetric setting such a
choice tends to spoil the running of the gauge couplings, while too
large a multiplicity of new states threatens perturbativity below the
ultraviolet scale at which the effective description is expected to
break down. The sector adopted here is consequently a compromise,
shaped jointly by anomaly cancellation, by the demand for improved
gauge-coupling evolution, and by the viability of the dark matter
candidate.

The general solution of the anomaly equations is described in
appendix~\ref{app:anomaly_constraints}, together with a second
representative charge assignment. The body of the paper is devoted to
one explicit model, analysed in full. This restriction is purely
practical: each distinct charge assignment, once supplemented by the
scalar fields and Yukawa couplings needed for symmetry breaking,
exotic-state decays and dark matter stability, amounts to an
independent renormalisable theory in its own right. Implementing any
one of these in
\SARAH~\cite{Staub:2008uz,Staub:2013tta,Goodsell:2014bna,Goodsell:2017pdq},
constructing the associated \SPheno~\cite{Porod:2003um,Porod:2011nf}
spectrum generator, and interfacing the output with
\MicrOMEGAs~\cite{Belanger:2010gh,Alguero:2023zol},
\SModelS~\cite{Kraml:2013mwa,Ambrogi:2018ujg,Alguero:2020grj,
Alguero:2021dig,Altakach:2024jwk} and
\HiggsTools~\cite{Bahl:2022igd} constitutes a self-contained
analysis, even when the workflow is automated through
\BSMArt~\cite{Goodsell:2023iac,Faraggi:2023jzm,BSMArtv2}. A
systematic survey of the full solution family is left for future
work.

The resulting construction is considerably more constrained than the
anomaly equations alone would suggest. Once every renormalisable
operator consistent with the symmetries is retained, the exotic states
are required to decay into lighter ones, stable coloured relics are
forbidden, and a neutral dark matter candidate must be present. These
requirements necessitate enlarging the minimal secluded fermion sector
by additional singlet scalars and Yukawa couplings. This scalar sector
is no mere addition: it is essential to the phenomenological coherence
of the anomaly-free completion, and it also gives rise to a light
pseudoscalar state \(A_1\), which opens additional annihilation channels
for the dark matter candidate and plays an important role in reducing
the relic density to the observed value across the viable parameter
space.

A discrete \(Z_2^{\rm DM}\) parity guarantees the stability of the
lightest neutral odd-sector state, which we identify as the dark
matter candidate. Its dark-matter phenomenology is related to, though
distinct from, standard \(Z^\prime\)-portal scenarios
\cite{Alves:2015pea,Dudas:2013sia,Lebedev:2014bba,Arcadi:2017kky}.
The relic density may be reduced through annihilation via the
baryonic \(Z^\prime\), through co-annihilation with nearby secluded
states \cite{Griest:1990kh}, or through annihilation into the light
pseudoscalar \(A_1\). Our numerical analysis further restricts the
parameter space: we concentrate on configurations in which part of the
secluded spectrum lies above the baryonic \(Z^\prime\). In this regime
the \(Z^\prime\) offers a characteristic resonant signature of the
model, while only part of the anomaly-free secluded spectrum would be
kinematically accessible at comparable energies; the remaining,
heavier states would call for exploration at still higher energies,
for instance at a Future Circular Collider. Within the viable region
we find a heavy \(Z^\prime\), a secluded sector containing even heavier
coloured and/or electrically charged states, a dark matter candidate
consistent with relic-density and direct-detection bounds, and
gauge-coupling trajectories that converge towards a common scale well
above the electroweak one.

We also examine a non-unified variant, in which the hidden gauge
coupling is treated as an independent parameter at the high scale.
This relaxes the usual bounds on the \(Z^\prime\) and permits spectra
in which the baryonic gauge boson sits at the TeV scale while some of
the anomaly-cancelling coloured states are pushed to considerably
higher masses. The phenomenology of this scenario bears some
resemblance to conventional heavy-\(Z^\prime\) and secluded-sector
dark matter constructions, but here the heavier states are not an
optional addition: they are demanded by the anomaly-free completion
itself.

Section~\ref{sec:construction} sets out the charge assignment, the
field content and the structure of the model.
Section~\ref{SEC:SARAH} describes the \SARAH implementation.
Section~\ref{sec:scan} explains the scan strategy and the selection
criteria. Section~\ref{sec:results} contains the phenomenological
analysis, and we conclude in section~\ref{sec:conclusion}.



\section{Construction of the model}
\label{sec:construction}

We now describe the construction of the anomaly-free baryonic completion
analysed in the numerical study. The new gauge factor is denoted by
\(\UB\), where \(\tB\) is an extended baryonic charge defined such that
\[
    \tB|_{\rm SM}=B ,
\]
so that the new symmetry coincides with ordinary baryon number on the
Standard Model (SM) fields. In the secluded sector, however, \(\tB\) is
not ordinary baryon number: its charge assignments are fixed jointly by
anomaly cancellation and by the renormalisable interactions required to
generate masses for the new states and to ensure a phenomenologically
viable theory.

The construction proceeds in two stages. We first identify a class of
anomaly-free charge assignments for a secluded fermion sector that is
vector-like under the SM gauge group but chiral under \(\UB\). This
fermion sector alone is not sufficient. Once one requires the exotic
states to decay, the dark matter candidate to remain stable, the scalar
symmetries to be broken appropriately, and unwanted relics to be absent,
the scalar sector and the Yukawa interactions must be enlarged
accordingly.

\begin{table*}[t]
\centering
\renewcommand{\arraystretch}{1.12}
\setlength{\tabcolsep}{7pt}
\begin{tabular}{llcccccc}
\toprule
Sector & Field & Generations & \(SU(3)_c\) & \(SU(2)_L\)
& \(U(1)_Y\) & \(\UB\) & \(\mathbb{Z}_2^{\rm DM}\) \\
\midrule
\multicolumn{8}{l}{\textit{Standard Model fields}} \\
\midrule
SM & \(Q_L\)        & \(3\) & \(3\)        & \(2\) & \(1/6\)  & \(1/3\)  & \(+1\) \\
   & \(u_R^c\)      & \(3\) & \(\bar 3\)  & \(1\) & \(-2/3\) & \(-1/3\) & \(+1\) \\
   & \(d_R^c\)      & \(3\) & \(\bar 3\)  & \(1\) & \(1/3\)  & \(-1/3\) & \(+1\) \\
   & \(L_L\)        & \(3\) & \(1\)        & \(2\) & \(-1/2\) & \(0\)    & \(+1\) \\
   & \(e_R^c\)      & \(3\) & \(1\)        & \(1\) & \(1\)    & \(0\)    & \(+1\) \\
   & \(\nu_R^c\)    & \(3\) & \(1\)        & \(1\) & \(0\)    & \(0\)    & \(+1\) \\
   & \(H\)          & \(1\) & \(1\)        & \(2\) & \(1/2\)  & \(0\)    & \(+1\) \\
\midrule
\multicolumn{8}{l}{\textit{Secluded fermions}} \\
\midrule
Secluded & \(\psi_L^L\)        & \(1\) & \(1\)       & \(2\) & \(-1/2\) & \(1/2\)  & \(-1\) \\
         & \((\psi_R^L)^c\)    & \(1\) & \(1\)       & \(2\) & \(1/2\)  & \(1\)    & \(-1\) \\
         & \(\psi_L^E\)        & \(1\) & \(1\)       & \(1\) & \(0\)    & \(-1/2\) & \(-1\) \\
         & \((\psi_R^E)^c\)    & \(1\) & \(1\)       & \(1\) & \(0\)    & \(-1\)   & \(-1\) \\
         & \(\psi_L^\nu\)      & \(1\) & \(1\)       & \(1\) & \(0\)    & \(-1/2\) & \(-1\) \\
         & \((\psi_R^\nu)^c\)  & \(1\) & \(1\)       & \(1\) & \(0\)    & \(-1\)   & \(-1\) \\
         & \(\psi_L^Q\)        & \(1\) & \(3\)       & \(2\) & \(1/6\)  & \(-2/3\) & \(-1\) \\
         & \((\psi_R^Q)^c\)    & \(1\) & \(\bar 3\) & \(2\) & \(-1/6\) & \(-5/6\) & \(-1\) \\
         & \(\psi_L^D\)        & \(2\) & \(3\)       & \(1\) & \(1/3\)  & \(2/3\)  & \(-1\) \\
         & \((\psi_R^D)^c\)    & \(2\) & \(\bar 3\) & \(1\) & \(-1/3\) & \(5/6\)  & \(-1\) \\
\midrule
\multicolumn{8}{l}{\textit{Secluded scalars}} \\
\midrule
Secluded & \(S_{3/2}\) & \(1\) & \(1\) & \(1\) & \(0\) & \(3/2\) & \(+1\) \\
         & \(S_1\)     & \(1\) & \(1\) & \(1\) & \(0\) & \(1\)   & \(-1\) \\
         & \(S_{1/2}\) & \(1\) & \(1\) & \(1\) & \(0\) & \(1/2\) & \(-1\) \\
         & \(X_1\)     & \(1\) & \(1\) & \(1\) & \(0\) & \(1\)   & \(+1\) \\
\bottomrule
\end{tabular}
\caption{\label{TAB:SARAHcontent}
Matter content of the model implemented in \SARAH, categorised by spin and
gauge representations.}
\end{table*}
\subsection{Charge assignments}
\label{subsec:charges_requirements}

Restricted to the Standard Model sector, the \(\UB\) charge coincides with
ordinary baryon number. In a left-chiral Weyl basis,
\[
    Q_L,\qquad u_R^c,\qquad d_R^c,\qquad L_L,\qquad e_R^c,\qquad \nu_R^c ,
\]
this means
\begin{equation}
    \tB(Q_L)=\frac13,\qquad
    \tB(u_R^c)=\tB(d_R^c)=-\frac13,
    \label{eq:SM_Btilde_quarks}
\end{equation}
and
\begin{equation}
    \tB(L_L)=\tB(e_R^c)=\tB(\nu_R^c)=\tB(H)=0.
    \label{eq:SM_Btilde_leptons}
\end{equation}
We include three right-handed neutrinos in the SM sector. They are neutral
under \(\UB\) in the baryonic benchmark, allowing the usual Dirac neutrino
Yukawa couplings. A Majorana mass term is also allowed and will be discussed
below.

Equivalently, in terms of physical right-handed quarks rather than their
left-chiral conjugates, all SM quarks carry baryon number \(1/3\), while
leptons and the Higgs are neutral. Since gauged baryon number is anomalous
with SM matter alone, additional chiral fermions are required.

The secluded fermion sector is chosen to be vector-like with respect to
\[
    SU(3)_c\times SU(2)_L\times U(1)_Y ,
\]
but chiral under \(\UB\). The labels
\[
    L,\qquad l,\qquad Q,\qquad D
\]
indicate the SM representation type of the corresponding secluded fields:
\(L\)-type fields are electroweak doublets, \(l\)-type fields are
colourless singlets, \(Q\)-type fields are colour-triplet electroweak doublets,
and \(D\)-type fields are colour-triplet electroweak singlets. These are
representation labels only; in particular, they should not be confused with a
conserved lepton number.

The secluded fermions acquire masses through Yukawa couplings to the
singlet scalar \(S\), whose vacuum expectation value breaks \(\UB\). If
\(q_X\) and \(\tilde q_X\) denote the \(\UB\) charges of a secluded left-chiral
fermion and its conjugate partner, gauge invariance of the corresponding Yukawa interaction requires
\begin{equation}
    \tilde q_X=-q_X+\epsilon_X q_S,
    \qquad X\in\{L,l,Q,D\}.
    \label{eq:main_sec_yukawa_constraint}
\end{equation}
The parameter \(q_S\) is the \(\UB\) charge of \(S\) and
\(\epsilon_X=\pm1\) specifies whether the coupling involves \(S\) or
\(S^\dagger\).

For the SM fields, gauge invariance of the Yukawa interactions implies
\begin{equation}
    \begin{split}
        z_e &= z_H - z_L, \quad z_\nu = -z_H - z_L, \\
        z_d &= z_H - z_Q, \quad z_u = -z_H - z_Q,
    \end{split}
    \label{eq:main_sm_yukawa_constraint}
\end{equation}
where \(z_Q,z_u,z_d,z_L,z_e,z_\nu,z_H\) are the \(\UB\) charges of
\(Q_L,u_R^c,d_R^c,L_L,e_R^c,\nu_R^c,H\), respectively, reducing to
Eqs.~\eqref{eq:SM_Btilde_quarks}--\eqref{eq:SM_Btilde_leptons} for the baryonic
assignment of the main numerical model.

Once the SM Yukawa relations are imposed, the non-trivial anomaly
coefficients satisfy
\begin{equation}
    \begin{split}
        t_{\tilde B} &= 0,
        \qquad
        t_{YY\tilde B} = -\frac12 t_2,
        \qquad
        t_{Y\tilde B\tilde B} = -2z_H t_2, \\
        t_{\tilde B\tilde B\tilde B} &= -6z_H^2 t_2,
        \qquad
        t_3 = 0,
    \end{split}
    \label{eq:main_sm_anomaly_relations}
\end{equation}
where \(t_2\) denotes the mixed \(SU(2)_L^2-\UB\) anomaly coefficient
and \(t_3\) the mixed \(SU(3)_c^2-\UB\) coefficient; full expressions and
conventions appear in appendix~\ref{app:anomaly_constraints}. After imposing the SM Yukawa relations, the only non-zero SM anomaly data relevant for
the baryonic branch are encoded in \(t_2\), to be cancelled by the secluded
sector.

To obtain a secluded sector with an SM-like organisation and vanishing
mixed gravitational and \(SU(3)_c^2-\UB\) anomaly coefficients, we impose
\begin{equation}
    N_l=2N_L,
    \quad \epsilon_l=-\epsilon_L,
    \quad
    N_D=2N_Q,
    \quad \epsilon_D=-\epsilon_Q ,
    \label{eq:main_NE_ND_conditions}
\end{equation}
where \(N_L,N_l,N_Q,N_D\) are the multiplicities of the corresponding
representations, and split the secluded singlets and colour triplets as
\begin{equation}
\begin{aligned}
    q_l^j &=
    \begin{cases}
        q_{E}, & j\le N_L,\\
        q_{\nu}, & j>N_L,
    \end{cases}
    &
    q_D^k &=
    \begin{cases}
        q_{d1}, & k\le N_Q,\\
        q_{d2}, & k>N_Q,
    \end{cases}
    \\
    q_Q^m&=q_Q,
    \qquad
    q_L^i=q_L,
\end{aligned}
\label{eq:main_charge_pattern}
\end{equation}
and similarly for the hypercharges. This notation anticipates the quark-like and lepton-like roles of the
corresponding secluded fields.

To keep the \(\tB\) charges rational and simplify the cubic anomaly
conditions, we further restrict to
\begin{equation}
    \begin{aligned}
        q_{d1} &= \pm(q_Q + z_H), \qquad & q_{d2} &= \pm(q_Q - z_H), \\
        q_{E}  &= \pm(q_L + z_H), \qquad & q_{\nu} &= \pm(q_L - z_H).
    \end{aligned}
    \label{eq:main_rationality}
\end{equation}
The derivation together with additional charge assignments is given in
appendix~\ref{app:anomaly_constraints}.

Anomaly cancellation alone still leaves a broad family of solutions. In the main text
we focus on one representative charge assignment satisfying additional
phenomenological requirements: the secluded sector must contain a neutral dark
matter candidate; the new matter should improve the high-scale evolution of the
SM gauge couplings; the hypercharges should remain close enough to SM values
that colour-singlet bound states have integer electric charge; and the \(\UB\)
charges should allow the scalar and Yukawa interactions needed for exotic-state
decays. A second representative charge assignment appears in
appendix~\ref{app:anomaly_constraints}, and a systematic analysis of the full
family of anomaly-free baryonic completions is left for future work.

The anomaly analysis thus determines a viable charge pattern for the fermion
sector, but not a complete phenomenological model. In particular, the minimal
construction containing only the secluded fermions and a single
\(\UB\)-breaking scalar would leave accidental global symmetries unbroken and
therefore lead to stable exotic states, which motivates the enlarged scalar and Yukawa sector described next.


\subsection{Model}
\label{subsec:model}

The phenomenological model contains the SM fields together with the secluded
fermions
\begin{equation}
    \psi_L,
    \qquad \psi_E,
    \qquad \psi_\nu,
    \qquad \psi_Q,
    \qquad \psi_D ,
\end{equation}
named according to their SM gauge representations. In addition to the SM Higgs
doublet \(H\), the scalar sector contains four SM-singlet fields,
\begin{equation}
    S_{3/2},
    \qquad S_1,
    \qquad S_{1/2},
    \qquad X_1 ,
\end{equation}
where the subscripts indicate their \(\tB\) charges,
\begin{equation}
    \begin{aligned}
        \tilde{B}(S_{3/2}) &= \frac32, \qquad & \tilde{B}(S_1) &= 1, \\
        \tilde{B}(S_{1/2}) &= \frac12, \qquad & \tilde{B}(X_1) &= 1.
    \end{aligned}
\end{equation}
Of these, \(S_{3/2}\) plays the leading role: it is the principal order
parameter for \(\UB\) breaking and generates the dominant secluded-fermion
masses. The remaining three singlets, \(S_1\), \(S_{1/2}\) and \(X_1\), play
a different role in the construction, to which we return shortly:
they break unwanted accidental symmetries, allow the exotic states to decay,
and shape the neutral sector from which the dark matter candidate emerges.

Stability of the dark matter candidate is guaranteed separately, by a
discrete parity \(Z_2^{\rm DM}\) assigned as
\begin{equation}
    \begin{aligned}
        \text{SM fields},\, H,\, S_{3/2},\, X_1 &: +1, \\
        \psi_L,\,\psi_E,\,\psi_\nu,\,\psi_Q,\,\psi_D,\, S_1,\, S_{1/2} &: -1.
    \end{aligned}
    \label{eq:Z2assign}
\end{equation}
The lightest neutral \(Z_2^{\rm DM}\)-odd state is then stable and is
identified as the dark matter candidate. The explicit charge assignment used
in the numerical implementation is given in Table~\ref{TAB:SARAHcontent},
which shows that \(\tB\) coincides with ordinary baryon number on the SM
fields but is an extended baryonic charge on the secluded sector.

The renormalisable Lagrangian is organised as
\begin{equation}
    \mathcal{L}
    =
    \mathcal{L}_{\rm kin}
    +\mathcal{L}_{\rm gauge}
    +\mathcal{L}_{\rm vev}
    +\mathcal{L}_{S}+
    \left(\mathcal{L}_{Y_1}
    +\mathcal{L}_{Y_2}
    +\mathcal{L}_{\rm br}
    +{\rm h.c.}\right)
    \label{eq:model_lagrangian_split}
\end{equation}
and it is worth setting out from the start what each piece is for, since the
logic of the construction rests on a division of labour between them.
\(\mathcal{L}_{\rm vev}\) and \(\mathcal{L}_{Y_1}\) together are responsible
for generating masses, both for the SM fields and for the leading
secluded-fermion states, through electroweak and \(\UB\) symmetry breaking.
Left on their own, however, these terms would leave the secluded sector with
an accidental symmetry that renders some of its states stable, including
coloured ones, which is phenomenologically unacceptable. The remaining terms play complementary roles. The interactions in
\(\mathcal{L}_S\) complete the singlet scalar potential and determine the
corresponding masses and quartic couplings, while
\(\mathcal{L}_{Y_2}\) and \(\mathcal{L}_{\rm br}\) remove the dangerous
accidental symmetries, open decay channels for the exotic states, and
determine the structure of the neutral sector relevant for dark matter.
We now examine each of these pieces in turn.

The Yukawa interactions responsible for the leading fermion masses are
contained in \(\mathcal{L}_{Y1}\),
\begin{equation}
\begin{aligned}
    \mathcal{L}_{Y1} = &
    - Y_u H Q_{L} u_R - Y_d H^\dagger Q_{L} d_R \\
    & - Y_e H^\dagger L_L e_R - Y_\nu H L_L\nu_R \\
    & - Y_{sq} \psi^Q_L \psi^Q_R S_{3/2} - Y_{sd} \psi^D_L \psi^D_R S_{3/2}^\dagger \\
    & - Y_{sl} \psi^L_L \psi^L_R S_{3/2}^\dagger - Y_{se} \psi^E_R \psi^E_L S_{3/2} \\
    & - Y_{s\nu} \psi^\nu_L \psi^\nu_R S_{3/2} - \frac{1}{2} M_R \nu_R \nu_R .
\end{aligned}
\label{eq:model_Y1}
\end{equation}
The first two lines are the usual SM Yukawa couplings, including the Dirac
neutrino coupling; a Majorana mass for \(\nu_R\) is allowed by the symmetries
and is discussed below. Taken alone, Eq.~\eqref{eq:model_Y1} leaves each
secluded Dirac pair carrying an independent conserved number, so the model
has an accidental symmetry
\begin{equation}
    U(1)_{Q'}\times U(1)_{D'}\times U(1)_{L'}
    \times U(1)_{E'}\times U(1)_{\nu'}
    \label{eq:model_accidental_symmetries}
\end{equation}
under which the lightest state in each sector is stable. Since some of these
states are coloured, this symmetry cannot survive if the model is to be
viable.

Before turning to the interactions that cure this, we record for completeness
the singlet bilinear and quartic terms in \(\mathcal{L}_S\),
\begin{equation}
\begin{aligned}
    \mathcal{L}_{S} = &
    - \mu_{S1}^2 S_{1} S_{1}^\dagger - \mu_{S12}^2 S_{1/2} S_{1/2}^\dagger \\
    & - \mu_X^2 X_{1} X_{1}^\dagger - \lambda_{S1} (S_{1} S_{1}^\dagger)^2 \\
    & - \lambda_{12} (S_{1/2} S_{1/2}^\dagger)^2 - \lambda_X (X_{1} X_{1}^\dagger)^2 \\
    & - \lambda_{112} (S_{1/2} S_{1/2}^\dagger)(S_{1} S_{1}^\dagger) \\
    & - \lambda_{X112} (S_{1/2} S_{1/2}^\dagger)(X_{1} X_{1}^\dagger) \\
    & - \lambda_{X1} (S_{3/2} S_{3/2}^\dagger)(S_{1} S_{1}^\dagger) \\
    & - \lambda_{X12} (S_{3/2} S_{3/2}^\dagger)(S_{1/2} S_{1/2}^\dagger) \\
    & - \lambda_{XX1} (S_{3/2} S_{3/2}^\dagger)(X_{1} X_{1}^\dagger).
\end{aligned}
\label{eq:model_LS}
\end{equation}
These terms fix the masses and quartic couplings of the odd and even
singlets, but they respect the accidental symmetries of
Eq.~\eqref{eq:model_accidental_symmetries} and so do nothing, by themselves,
to solve the problem just identified.

The accidental symmetries are broken by two additional sets of interactions. The Yukawa couplings in
\(\mathcal{L}_{Y_2}\),
\begin{equation}
\begin{aligned}
    \mathcal{L}_{Y2} = &
    - Y_{d12} d_R \psi^D_R S_{1/2}^\dagger - Y_{q12} Q_{L} \psi^Q_R S_{1/2} \\
    & - Y_{l1} L_L \psi^L_R S_{1}^\dagger - Y_{e1} \psi^E_R S_{1} \nu_R \\
    & - Y_{\nu1} S_{1} \psi^\nu_R \nu_R - Y_{e12} \psi^E_L S_{1/2} \nu_R \\
    & - Y_{\nu12} S_{1/2} \psi^\nu_L \nu_R - Y_{Xee} (\psi^E_L)^2 X_{1} \\
    & - Y_{Xe\nu} \psi^E_L \psi^\nu_L X_{1} - Y_{X\nu\nu} (\psi^\nu_L)^2 X_{1} ,
\end{aligned}
\label{eq:model_Y2}
\end{equation}
connect the secluded fermions to the SM and to one another through \(S_1\),
\(S_{1/2}\) and \(X_1\), while the scalar trilinear and quartic
interactions in \(\mathcal{L}_{\rm br}\),
\begin{equation}
\begin{aligned}
    \mathcal{L}_{\rm br} = &
    - k_{X1} (S_{1/2})^2 X_{1}^\dagger
    - k_{X} S_{3/2} S_{1}^\dagger S_{1/2}^\dagger \\
    & - \lambda_{SX11} (S_1)^2 (X_1^\dagger)^2 \\
    & - \lambda_{A} S_{1/2} S_{3/2} S_{1}^\dagger X_{1}^\dagger .
\end{aligned}
\label{eq:model_Lbr}
\end{equation}
link the singlets among themselves. Together, \(\mathcal{L}_{Y_2}\) and
\(\mathcal{L}_{\rm br}\) remove the accidental continuous symmetries of
Eq.~\eqref{eq:model_accidental_symmetries} that would otherwise protect
the exotic states. This has three important
phenomenological consequences: the heavier exotic states can now decay
into lighter secluded states and SM particles, no coloured relic remains
stable, and the neutral \(Z_2^{\rm DM}\)-odd sector acquires the non-trivial
structure from which the dark matter candidate is drawn. The same interaction network controls the explicit breaking of the
accidental \(X_1\)-phase symmetry and hence the mass of the light
pseudoscalar discussed in Section~\ref{subsec:fate_A1}.

The remaining piece, \(\mathcal{L}_{\rm vev}\),
\begin{equation}
\begin{aligned}
    \mathcal{L}_{\rm vev} = &
    - \mu_H^2 H^\dagger H - \mu_{S32}^2 S_{3/2}^\dagger S_{3/2} \\
    & - \lambda_H (H^\dagger H)^2 - \lambda_S (S_{3/2}^\dagger S_{3/2})^2 \\
    & - \lambda_{HS} (S_{3/2}^\dagger S_{3/2}) (H^\dagger H),
\end{aligned}
\label{eq:model_Lvev}
\end{equation}
generates the vacuum expectation values responsible for electroweak and
\(\UB\) symmetry breaking, and hence for the fermion and gauge-boson masses.
The interactions involving the extended scalar sector therefore play two
distinct roles in the model. The terms in \(\mathcal{L}_{\rm vev}\) set the
scale of symmetry breaking and the bulk of the mass spectrum, while those in
\(\mathcal{L}_S\), \(\mathcal{L}_{Y_2}\), and
\(\mathcal{L}_{\rm br}\) determine the accidental-symmetry structure,
the allowed decays of the exotic states, and the origin of the light
pseudoscalar. Both functions are required in the phenomenological model
considered here.

The symmetry-breaking pattern is
\begin{equation}
    \langle H^0\rangle=\frac{v}{\sqrt2},
    \quad
    \langle S_{3/2}\rangle=\frac{v_S}{\sqrt2},
    \quad
    v_S\gg v ,
    \label{eq:model_vevs}
\end{equation}
where \(v\) is the electroweak scale and \(v_S\) sets the scale of \(\UB\)
breaking. The baryonic gauge boson acquires a mass
\begin{equation}
    m_{Z^\prime}\simeq \frac32\,g_X v_S ,
    \label{eq:model_mZp}
\end{equation}
up to kinetic and mass mixing effects, where \(g_X\) is the \(\UB\) gauge
coupling. In the numerical implementation, the field \(X_1\) may acquire a
small loop-induced vacuum expectation value. The associated pseudoscalar is
typically much lighter than the main secluded-sector scale in the viable
benchmarks and plays an important role in some of the dark matter regions.

After symmetry breaking, the secluded fermion spectrum contains one up-type
quark-like state, three down-type quark-like states, one charged lepton-like
state, and six neutral states,
\begin{equation}
    \tilde u,
    \quad
    \tilde d_i\ (i=1,2,3),
    \quad
    \tilde e,
    \quad
    \tilde\nu_a\ (a=1,\ldots,6).
    \label{eq:model_fermion_spectrum}
\end{equation}
The neutral states arise from the mixing of the neutral components of
\(\psi_L\), \(\psi_E\), and \(\psi_\nu\), and the lightest is the dark matter
candidate.

\subsection{Lepton number}
\label{subsec:lepton_number}

The new gauge symmetry is baryonic rather than leptonic: the SM leptons are
neutral under \(\UB\), so lepton number is not protected by the gauge
symmetry. In the visible sector alone, the Yukawa interaction
\[
    -Y_\nu H L_L\nu_R+{\rm h.c.}
\]
is compatible with the usual lepton-number assignment of the SM leptons and
generates Dirac neutrino masses. In the full model, however, lepton number is
not an exact symmetry.

For the benchmark charge assignment in
Table~\ref{TAB:SARAHcontent}, \(\nu_R\) is neutral under \(\UB\) and even
under \(Z_2^{\rm DM}\). The Majorana mass term
\begin{equation}
    -\frac12\,M_R\,\nu_R\nu_R+{\rm h.c.}
    \label{eq:model_MR}
\end{equation}
is therefore allowed. If present, it violates lepton number by two units and
gives rise to the usual seesaw structure for the visible neutrinos.

Even if Eq.~\eqref{eq:model_MR} is absent at tree level, the
renormalisable interactions of the full model do not preserve any simple
extension of the SM lepton number. This can already be seen from the three
interactions
\begin{equation}
    \psi_L^E S_{1/2}\nu_R,
    \qquad
    \psi_L^E\psi_L^E X_1,
    \qquad
    X_1^\dagger S_{1/2}^2 .
    \label{eq:model_L_breaking_terms}
\end{equation}
Suppose one assigns a global lepton number \(L\) extending the SM
assignment, with
\[
    L(L_L)=L(\nu_R)=1,
    \qquad
    L(H)=0.
\]
Invariance of the first two interactions in
Eq.~\eqref{eq:model_L_breaking_terms} requires
\begin{equation}
    L(\psi_L^E)+L(S_{1/2})+1=0,
    \qquad
    2L(\psi_L^E)+L(X_1)=0,
\end{equation}
which implies
\begin{equation}
    L(X_1)=2L(S_{1/2})+2.
\end{equation}
The scalar interaction \(X_1^\dagger S_{1/2}^2\), however, requires
\begin{equation}
    -L(X_1)+2L(S_{1/2})=0,
\end{equation}
or equivalently
\begin{equation}
    L(X_1)=2L(S_{1/2}).
\end{equation}
The two conditions are incompatible. Consequently, no conserved global
lepton number can simultaneously extend the SM assignment and leave all the
required interactions invariant.

Once the Yukawa interactions linking \(\nu_R\) to the secluded neutral
fermions and the scalar interaction \(X_1^\dagger S_{1/2}^2\) are present,
lepton number is necessarily violated by two units. The same conclusion
applies to the analogous couplings involving \(\psi_L^\nu\). The model
therefore naturally accommodates Majorana neutrino masses and loop
corrections to the neutral-fermion mass matrix. A detailed study of the
loop-corrected neutrino sector is left for future work.

The exact \(Z_2^{\rm DM}\) parity nevertheless remains essential. Since the
odd scalars do not acquire vacuum expectation values,
\begin{equation}
    \langle S_1\rangle=\langle S_{1/2}\rangle=0,
    \label{eq:model_odd_scalar_vevs}
\end{equation}
the visible neutrinos do not mix at tree level with the
\(Z_2^{\rm DM}\)-odd neutral fermions. Dark matter stability is therefore
preserved even though lepton number is not an exact symmetry.

\subsection{Proton stability}

Since \(U(1)_{\tilde B}\) acts as baryon number on the SM fields, the usual
dimension-six proton-decay operators built solely from SM fields carry
non-zero \(\tB\) charge. In the baryonic normalisation,
\begin{equation}
    \begin{aligned}
        \tilde B(QQQL) &= 1, \\
        \tilde B(u^cu^cd^ce^c) &= -1, \\
        \tilde B(u^cd^cd^c\nu^c) &= -1 .
    \end{aligned}
\end{equation}
These operators are therefore absent before \(U(1)_{\tilde B}\) breaking.
They may nevertheless appear at higher dimension once dressed by singlet
fields. Since \(X_1\) is even under \(Z_2^{\rm DM}\) and carries
\(\tilde B(X_1)=1\), the operators
\[
    \frac{X_1^\dagger}{\Lambda^3}QQQL,
    \qquad
    \frac{X_1}{\Lambda^3}u^cu^cd^ce^c,
    \qquad
    \frac{X_1}{\Lambda^3}u^cd^cd^c\nu^c
\]
are allowed by the low-energy symmetries. After
\(\langle X_1\rangle=v_X/\sqrt2\), they induce effective dimension-six
operators suppressed by
\[
    \frac{1}{\Lambda_{\rm eff}^2}
    =
    \frac{v_X}{\sqrt2\,\Lambda^3},
    \qquad
    \Lambda_{\rm eff}
    =
    \left(\frac{\sqrt2\,\Lambda^3}{v_X}\right)^{1/2}.
\]
For generic flavour coefficients of order unity involving first-generation
fields, current proton-decay limits require
\(\Lambda_{\rm eff}\gtrsim10^{15}\text{--}10^{16}\,\mathrm{GeV}\). With
\(\Lambda\sim10^{16}\,\mathrm{GeV}\) and
\(v_X\lesssim\mathrm{TeV}\), one obtains
\[
    \Lambda_{\rm eff}\gtrsim10^{22}\,\mathrm{GeV},
\]
comfortably above the experimental bound. Proton stability therefore
requires either a sufficiently high ultraviolet scale, a sufficiently small
\(v_X\), or an additional selection rule suppressing these operators. In the
phenomenological analysis we restrict ourselves to parameter regions where
this condition is satisfied.

The same symmetry assignments also constrain charged-lepton flavour
violation. Since
\(\langle S_1\rangle=\langle S_{1/2}\rangle=0\), the SM charged leptons do
not mix at tree level with the \(Z_2^{\rm DM}\)-odd secluded leptons.
Loop-induced charged-lepton flavour violation can nevertheless arise from
the Yukawa couplings connecting \(L_L\) to the secluded lepton-like states,
with amplitudes controlled by flavour off-diagonal combinations such as
\[
    (Y_{l1}^\dagger Y_{l1})_{\beta\alpha}.
\]
In the benchmark scan these Yukawa matrices are taken to be flavour
aligned, so the off-diagonal combinations vanish or are sufficiently
suppressed. More general flavour structures require a dedicated analysis.


\subsection{The pseudoscalar \texorpdfstring{$A_1$}{A1}}
\label{subsec:fate_A1}

The scalar sector possesses an accidental global symmetry associated with the
phase of the field \(X_1\),
\begin{equation}
    U(1)_X:\qquad
    X_1\rightarrow e^{i\alpha}X_1,
    \qquad
    \text{all other fields fixed},
    \label{eq:A1_U1X}
\end{equation}
when the interactions carrying a net \(X_1\)-number are absent.
Writing
\[
    X_1=\frac{1}{\sqrt2}(v_X+x_1+iA_1),
\]
the field \(A_1\) denotes the physical pseudoscalar predominantly aligned with
the phase of \(X_1\), after removing the Goldstone bosons associated with the
electroweak and \(\UB\) symmetry breaking.

The interactions that explicitly violate the \(U(1)_X\) phase symmetry are
those appearing in
Eqs.~\eqref{eq:model_Y2} and~\eqref{eq:model_Lbr}.  Taken individually,
these interactions do not necessarily remove every residual continuous
rephasing symmetry.  The explicit breaking therefore results from the combined
interaction structure of the scalar and fermion sectors rather than from any
single coupling.

The numerical implementation allows two qualitatively different situations.
If \(X_1\) does not acquire a vacuum expectation value at tree level,
the tree-level potential depends only on \(|X_1|^2\), and the CP-even and
CP-odd components of \(X_1\) are degenerate at tree level. Radiative effects
generate a small vacuum expectation value through the \(X_1\) tadpole and lift
this degeneracy.

Alternatively, \(X_1\) may acquire a tree-level vacuum expectation value,
spontaneously breaking the accidental \(U(1)_X\) symmetry. In that case,
\(A_1\) is the corresponding Goldstone boson at tree level, and its mass is
generated only through the interactions that explicitly break the accidental
symmetry.

The benchmark points presented in this work belong to the second class.
Consequently, they contain a light pseudoscalar whose mass originates from the
explicit breaking of the accidental \(U(1)_X\) symmetry and which plays an
important role in the dark matter phenomenology discussed below.




\subsection{Phenomenological regime of interest}
\label{subsec:pheno_regime}

The construction itself does not require a hierarchical spectrum beyond the
Standard Model. In the phenomenological analysis, however, we concentrate on
the region in which the baryonic \(Z^\prime\) lies below part of the secluded
sector,
\begin{equation}
    m_{Z^\prime}\gtrsim {\rm few\ TeV},
    \qquad
    m_{\rm heavy\ sec}>m_{Z^\prime},
    \qquad
    m_{\rm DM}<m_{Z^\prime}.
    \label{eq:model_pheno_regime}
\end{equation}
In this regime the baryonic \(Z^\prime\) provides the first characteristic
signature of the model, whereas only part of the secluded spectrum is
kinematically accessible at comparable energies. The remaining secluded states
would require higher-energy exploration. Realising this spectrum requires a
non-trivial interplay among the Yukawa couplings: some must be sufficiently
large to raise part of the secluded spectrum above the \(Z^\prime\), whereas
others must remain compatible with a neutral dark matter candidate and with
perturbative couplings up to high scales.

The benchmark points retained in the scan all contain a comparatively light,
predominantly singlet-like pseudoscalar \(A_1\). We deliberately restrict the
present study to this region of parameter space. For heavier \(A_1\), the
fixed-order calculation performed by \SPheno develops sizeable radiative
corrections to the pseudoscalar pole mass, making the perturbative prediction
progressively less reliable. A dedicated resummation or effective-potential
analysis would then be required, and we leave this regime for future work.

This restriction is consistent with current Higgs phenomenology. Throughout
the benchmark points, the mixing of \(A_1\) with the SM-like Higgs sector is
negligible, and the corresponding contribution to the Higgs width is typically
of order
\[
    \Gamma(h\rightarrow A_1A_1)\sim10^{-11}\ {\rm GeV},
\]
far below the present sensitivity of Higgs-width and exotic-Higgs-decay
measurements. In the phenomenological scenarios studied here, \(A_1\) therefore
acts essentially as a secluded-sector state whose main impact is to provide an
additional annihilation channel for the dark matter candidate.



\section{Implementation in \SARAH}
\label{SEC:SARAH}

A complete phenomenological analysis of this model requires the simultaneous
treatment of several non-trivial features: a large mixed scalar sector, six
neutral secluded fermions, loop-corrected masses, and the constraints from
collider, Higgs, and dark matter observables. We therefore implement the model
of Table~\ref{TAB:SARAHcontent} in \SARAH, which generates the Fortran code
interfaced with \SPheno for the spectrum calculation. The resulting spectrum
and vertices are subsequently passed to the phenomenological tools described
below, while the full workflow is automated through \BSMArt.

The exact \(Z_2^{\rm DM}\) parity separates, apart from the even singlets, the
SM and secluded sectors and guarantees the stability of the lightest neutral
odd state. After symmetry breaking, the six neutral secluded fermions mix, and
the lightest viable Majorana eigenstate is identified as the dark matter
candidate.

The fields \(\psi^{E}_{L/R}\) and \(\psi^{\nu}_{L/R}\) have identical gauge
quantum numbers. They are nevertheless implemented as distinct fields, both
to retain the notation of Section~\ref{subsec:model} and to keep their
independent couplings explicit in the \SARAH model file. The odd scalars
\(S_1\) and \(S_{1/2}\) couple to both the visible and secluded sectors,
opening decay channels for the heavier secluded states. The even scalar
\(X_1\) enters the neutral-fermion sector and participates in the explicit
breaking of the accidental global symmetries. As discussed in
Section~\ref{subsec:model}, these additional fields are essential for avoiding
stable coloured exotic states and for obtaining the neutral-sector structure
relevant for dark matter.

The dictionary between the notation used in the paper and the corresponding
parameter names in the \SARAH model is given in
Appendix~\ref{app:dictionary}.

The scalar and neutral-fermion sectors are coupled through numerous
interactions. Electroweak and \(U(1)_{\tilde B}\) breaking are driven
primarily by the vacuum expectation values of \(H\) and \(S_{3/2}\).
Interactions involving \(X_1\), such as \(Y_{X\nu\nu}\), generate an
\(X_1\) tadpole at one loop and therefore induce a small vacuum expectation
value for this field in the numerical calculation. This radiatively induced
expectation value is included consistently in the determination of the scalar
spectrum and of the neutral-fermion mixing.

\subsection{Scalar sector}

The scalar sector splits into \(Z_2^{\rm DM}\)-even and \(Z_2^{\rm DM}\)-odd
fields. The even fields are
\begin{equation}
    H,\qquad S_{3/2},\qquad X_1 ,
\end{equation}
whereas \(S_1\) and \(S_{1/2}\) are odd. The odd scalars do not acquire
vacuum expectation values. We write
\begin{equation}
    S_1 = \frac{1}{\sqrt2}(s_1+i a_1),
    \qquad
    S_{1/2} = \frac{1}{\sqrt2}(s_{1/2}+i a_{1/2}) .
\end{equation}
The scalar and pseudoscalar components mix among themselves, giving two
\(Z_2^{\rm DM}\)-odd scalar mass eigenstates \(\sigma_i\) and two odd
pseudoscalar mass eigenstates \(\phi_i\), with \(i=1,2\). Their masses are
determined primarily by the input parameters \(\mu_{S1}^2\) and
\(\mu_{S12}^2\), which are varied in the scan, together with additional
contributions proportional to \(v_S\).

The even sector is responsible for electroweak and
\(U(1)_{\tilde B}\) symmetry breaking. The Goldstone bosons associated with
\(H\) and \(S_{3/2}\) are eaten by the \(Z\) and \(Z^\prime\), leaving a
SM-like Higgs boson, a predominantly-\(S_{3/2}\) scalar associated with
\(U(1)_{\tilde B}\) breaking, and the two real components of \(X_1\). We
write
\begin{equation}
    X_1 = \frac{1}{\sqrt2}(v_X+x_1+iA_1).
\end{equation}

The expectation value \(v_X\) is generally expected to be non-zero, since no
symmetry forbids an \(X_1\) tadpole (in contrast, the \(Z_2^{\rm DM}\) parity
prevents analogous terms for the odd scalars). The scalar potential may be
written schematically as
\begin{equation}
\begin{aligned}
    V \simeq\;&
    T_X (v_X+x_1) \\
    &+\frac12\mu_X^2\left[(v_X+x_1)^2+A_1^2\right] \\
    &+\frac14\lambda_X
    \left[(v_X+x_1)^2+A_1^2\right]^2 ,
\end{aligned}
\end{equation}
where \(T_X\) denotes the loop-induced tadpole. The minimisation condition is
\begin{equation}
    0=T_X+\mu_X^2v_X+\lambda_Xv_X^3.
\end{equation}
For a sufficiently small tadpole, this naturally gives
\(v_X\ll v_S\), provided \(\lambda_X\) is not too small.

In the present implementation, \(v_X\) is treated as an input parameter,
while the tadpole equation is solved for \(\mu_X^2\). This choice is
convenient for the \SARAH/\SPheno calculation of the scalar and
pseudoscalar pole masses.

With this parametrisation, however, the numerical implementation does not
distinguish between two physically different situations. One possibility is
that the accidental \(U(1)_X\) symmetry is already spontaneously broken at
tree level through a tachyonic value of \(\mu_X^2\). The other is that the
tree-level potential preserves the accidental symmetry and that the non-zero
value of \(v_X\) originates from the radiatively generated tadpole. Once
\(v_X\) is specified as an input, both possibilities correspond to the same
low-energy parametrisation, and the present implementation does not identify
which underlying realisation is selected.

The benchmark points considered in this work all contain a light
pseudoscalar \(A_1\), whose properties are discussed in
Section~\ref{subsec:fate_A1}. In the region of parameter space explored here,
the one-loop and two-loop corrections to its pole mass can be numerically
comparable, making higher-order corrections important for a reliable
prediction.

For the \(Z_2^{\rm DM}\)-even scalar masses, labelled
\(h_1,h_2,h_3\), the tree-level mass matrix is
\begin{equation}
    (\mathcal{M}_h^2)^{\rm tree} =
    \begin{pmatrix}
        2\lambda_Hv^2 &
        \lambda_{HS}vv_S &
        0 \\[1mm]
        \lambda_{HS}vv_S &
        2\lambda_Sv_S^2 &
        \lambda_{XX1}v_Xv_S \\[1mm]
        0 &
        \lambda_{XX1}v_Xv_S &
        2\lambda_Xv_X^2
    \end{pmatrix}.
\end{equation}
The lightest eigenstate is identified with the observed 125 GeV Higgs boson.

\subsection{\texorpdfstring{The $Z^\prime$ boson}{The Z-prime boson}}
\label{subsec:Zprime}

The mass of the baryonic \(Z^\prime\) is approximately
\begin{equation}
    m_{Z^\prime}\simeq \frac32\,g_Xv_S .
\end{equation}
The numerical calculation includes the full kinetic and mass mixing.
Throughout the phenomenological analysis we focus on a heavy baryonic
gauge boson, corresponding to values of \(v_S\) large enough to satisfy
current experimental constraints from \(Z^\prime\) searches.

\subsection{Fermions}

The remaining spectrum consists of the SM fermions together with the secluded
fermion sector. The secluded fermion masses are determined by \(v_S\) and the
corresponding Yukawa couplings. After electroweak and
\(U(1)_{\tilde B}\) symmetry breaking, the physical spectrum contains
\begin{itemize}
    \item three secluded down-type quarks, \(\tilde d_i\), originating from
          \(\psi_{L/R}^D\) and \(\psi_{L/R}^Q\);
    \item one secluded up-type quark, \(\tilde u\), originating from
          \(\psi_{L/R}^Q\);
    \item one secluded charged lepton, \(\tilde e\), originating from
          \(\psi_{L/R}^L\);
    \item six secluded neutral fermions, originating from
          \(\psi_{L/R}^L\), \(\psi_{L/R}^E\), and
          \(\psi_{L/R}^\nu\).
\end{itemize}

At tree level, an approximate symmetry causes the neutral components of
\(\psi_{L/R}^L\) to form a pseudo-Dirac pair, while the remaining four
neutral states are Majorana fermions, denoted by \(\tilde\nu_i\).

The phenomenological analysis focuses on spectra in which at least one
secluded quark is heavier than the \(Z^\prime\). This requires some of the
secluded Yukawa couplings at the \(U(1)_{\tilde B}\)-breaking scale to be
larger than the gauge coupling \(g_X\).

Two limiting cases of the neutrino sector are particularly useful.

In the first, the right-handed neutrino Majorana mass \(M_R\) is non-zero.
Together with the neutrino-portal couplings linking \(\nu_R\) to the secluded
neutral fermions, it breaks lepton number and generates loop-induced
splittings in the \(\psi_{L/R}^L\) sector. For sufficiently large \(M_R\),
the visible neutrino masses are dominated by the seesaw mechanism, allowing
the Dirac neutrino Yukawa coupling \(Y_\nu\) to be sizeable. Such splittings
can modify the dark matter phenomenology by preventing the heavier neutral
states from becoming long-lived. On the other hand, a large value of
\(M_R\) suppresses neutrino-mediated decays of the secluded states and may
itself lead to long-lived particles. This case has been implemented in
\SARAH, but a detailed phenomenological study is left for future work.

The benchmark scans presented here correspond to the second limit. We set
\(M_R=0\) and take \(Y_\nu\) to be negligible; in the \SARAH
implementation, it is set identically to zero. Within this scan setup, this
restores an exact anomalous lepton-number-like symmetry. This is a technical
simplification, corresponding to a particular limit of the full model,
whose complete interaction structure does not preserve lepton number, as
discussed in Section~\ref{subsec:lepton_number}.

In this limit, the neutral components of \(\psi_{L/R}^L\) form a Dirac
secluded neutrino, denoted by \(\tilde\nu_D\), while the remaining neutral
states are labelled \(\tilde\nu_i\), \(i=1,\ldots,4\). In the flavour basis
\((\psi^E_L,\psi^\nu_L,(\psi^E_R)^c,(\psi^\nu_R)^c)\), the mass matrix is
\begin{align}
M_{\tilde{\nu}}=
\begin{pmatrix}
\sqrt{2}v_XY_{Xee} &
\dfrac{v_XY_{Xev}}{\sqrt2} &
\dfrac{v_SY_{se}}{\sqrt2} &
0\\
\dfrac{v_XY_{Xev}}{\sqrt2} &
\sqrt{2}v_XY_{Xvv} &
0 &
\dfrac{v_SY_{sv}}{\sqrt2}\\
\dfrac{v_SY_{se}}{\sqrt2} &
0 &
0 &
0\\
0 &
\dfrac{v_SY_{sv}}{\sqrt2} &
0 &
0
\end{pmatrix}.
\end{align}

The Dirac secluded neutrino has mass
\[
m_{\tilde\nu_D}=\frac{Y_{sl}v_S}{\sqrt2}.
\]
It is not a viable dark matter candidate because it would be excluded by
direct-detection constraints. Throughout the viable benchmark points, the
dark matter candidate is instead the lightest Majorana state,
\(\tilde\nu_1\).

\subsection{Exploration of the parameter space}
\label{sec:scan}

The implementation in \SARAH generates a spectrum calculator based on
\SPheno. The resulting spectrum is then passed to the phenomenological
tools, while the code generation, interfaces, and parameter scans are
handled automatically by \BSMArt version~2 \cite{BSMArtv2}.

The model contains a large number of dimensionless couplings. Among the
dimensionful parameters, the vacuum minimisation conditions determine
three mass-squared parameters (\(\mu_H^2\), \(\mu_S^2\), and
\(\mu_X^2\)) once the vacuum expectation values \(v_H\), \(v_S\), and
\(v_X\) are specified. The remaining dimensionful free parameters are
therefore \(\mu_{S1}^2\), \(\mu_{S12}^2\), \(k_X\), and \(k_{X1}\).
The SM parameters are fixed at low energies. \SARAH determines the gauge
coupling unification scale, typically slightly above
\(10^{16}\,\mathrm{GeV}\), where the remaining new couplings and masses
are specified. The Higgs quartic coupling \(\lambda_H\) is adjusted
using the fitting functionality of \BSMArt to reproduce the observed
Higgs mass of \(125\,\mathrm{GeV}\).

The remaining free parameters are listed in
Table~\ref{TAB:scanranges}. This leaves a 17-dimensional parameter
space, too large for either a regular grid scan or a naive
Markov-chain Monte Carlo to explore efficiently.

\begin{table}[t]
\centering
\renewcommand{\arraystretch}{1.12}
\setlength{\tabcolsep}{6pt}
\begin{tabular}{lcc}
\toprule
Coupling & Range & Prior \\
\midrule
\(Y_{sd}\) & $[-0.9,\,0.9]$ & linear \\
\(Y_{sq}\) & $[-0.9,\,0.9]$ & linear \\
\(Y_{su}\) & $[-0.9,\,0.9]$ & linear \\
\(Y_{s\nu}\) & $[-0.7,\,0.7]$ & linear \\
\(Y_{se}\) & $[-0.7,\,0.7]$ & linear \\
\(Y_{sl}\) & $[-0.7,\,0.7]$ & linear \\
\(\lambda_{HS}\) & $[0,\,1.0]$ & linear \\
\(\lambda_{S}\) & $[0.0,\,1.5]$ & linear \\
\(Y_{Xee}\) $=$ \(Y_{X\nu\nu}\) & $[-0.9,\,0.9]$ & linear \\
\(\lambda_X\) & $[0.0,\,1.5]$ & linear \\
$v_S$ & $[10^4,\,5\times10^5]$ & linear \\
$v_X$ & $[1,\,10^4]$ & linear \\
\(y_{off}\) & $[-0.9,\,0.9]$ & linear \\
\(k_X\) & $[10,\,10^4]$ & log \\
\(k_{X1}\) & $[10,\,10^4]$ & log \\
\(m_{S1}\) & $[10^3,\,10^5]$ & log \\
\(\mu_{S12}\) & $[10^3,\,10^5]$ & log \\
\bottomrule
\end{tabular}
\caption{\label{TAB:scanranges}
Ranges and priors for the scan parameters. We scan over \(m_{S1}\) and
\(m_{S12}\), and set \(\mu_{S1}^2=m_{S1}^2\) and
\(\mu_{S12}^2=m_{S12}^2\). We fix \(Y_{Xee}=Y_{Xvv}\)
and take \(Y_{Xev}=Y_{Xve}=0\). A common visible--secluded Yukawa
parameter \(y_{off}\) is used:
\(Y_{d12}^{ij}=y_{off}\,\delta^{ij}\) and
\(Y_{e1}^i=Y_{e12}^i=Y_{l1}^i
=Y_{q12}^i=Y_{\nu1}^i=Y_{\nu12}^i=y_{off}\)
for \(i=1,2,3\). The quartic couplings \(\lambda_{12}\), \(\lambda_{112}\),
\(\lambda_{S1}\), \(\lambda_{X1}\), \(\lambda_{X12}\),
\(\lambda_{XX1}\), \(\lambda_{X112}\), and \(\lambda_{SX11}\) are all fixed
to \(0.1\), and \(\lambda_{A}=0.4\). The non-zero value of
\(\lambda_{SX11}\) is consistent with Section~\ref{subsec:fate_A1}, where it
appears as one of the spurions entering the collective breaking of the \(X_1\)
phase symmetry. This leaves 17 scanned parameters.}
\end{table}

We therefore adopt a two-step procedure. First, we perform a
\texttt{CMAES} scan
\cite{CMAES-original,nomura2026cmaessimplepracticalpython,deSouza:2022uhk}
using \BSMArt~v2
\cite{Goodsell:2023iac,Faraggi:2023jzm,BSMArtv2}. This evolutionary
algorithm optimises a likelihood function and is both efficient and
robust---an important property, since many parameter points fail to
produce a valid spectrum with the current toolchain. We use linear
priors for most parameters and logarithmic priors for those spanning
several orders of magnitude. The likelihood is constructed from the
following observables: a preference for \(Z^\prime\) masses above the
current limit of \(5\,\mathrm{TeV}\) (already largely enforced by the
chosen range of \(v_S\)); positive mass splittings between the
lightest up-type, down-type, and charged-lepton-like secluded states
and the dark matter candidate; a dark matter relic abundance
\(\Omega h^2<0.112\); the direct-detection \(p\)-value returned by
\MicrOMEGAs; and the \SModelS upper-limit cross-section ratio required
to remain below \(0.8\). The toolchain (\SPheno, \MicrOMEGAs, and
\SModelS) is executed sequentially. If a validity criterion fails at
any stage, the remaining tools are skipped and the corresponding
likelihood terms are assigned their maximum penalty values. The
algorithm therefore retains information on excluded points without
wasting computational time. The \BSMArt templates used for our scans
are available at
\url{https://github.com/BSMArt-HEP/examples}.

The optimal points obtained from the \texttt{CMAES} scan are then used
as seeds for an affine-invariant Markov-chain Monte Carlo (MCMC)
based on the Goodman--Weare stretch-move algorithm
\cite{Goodman:2010dyf}, using the same likelihood function. This
produces a sample distributed according to the posterior,
concentrating around the regions of highest likelihood while still
exploring lower-likelihood regions in a controlled manner. The
affine-invariant algorithm is particularly well suited to
high-dimensional parameter spaces, since it proposes new points along
lines connecting a fixed number of walkers rather than drawing from an
\(n\)-dimensional Gaussian distribution.

From the resulting sample we retain the points satisfying all
selection criteria and subsequently check the Higgs constraints using
\HiggsTools. Since the light scalar states are predominantly SM
singlets with negligible mixing with the Higgs sector, they are not
constrained by current Higgs measurements.

\subsection{Analysis}
\label{sec:results}

\begin{figure*}[p]
    \centering
    \includegraphics[width=0.72\textwidth]{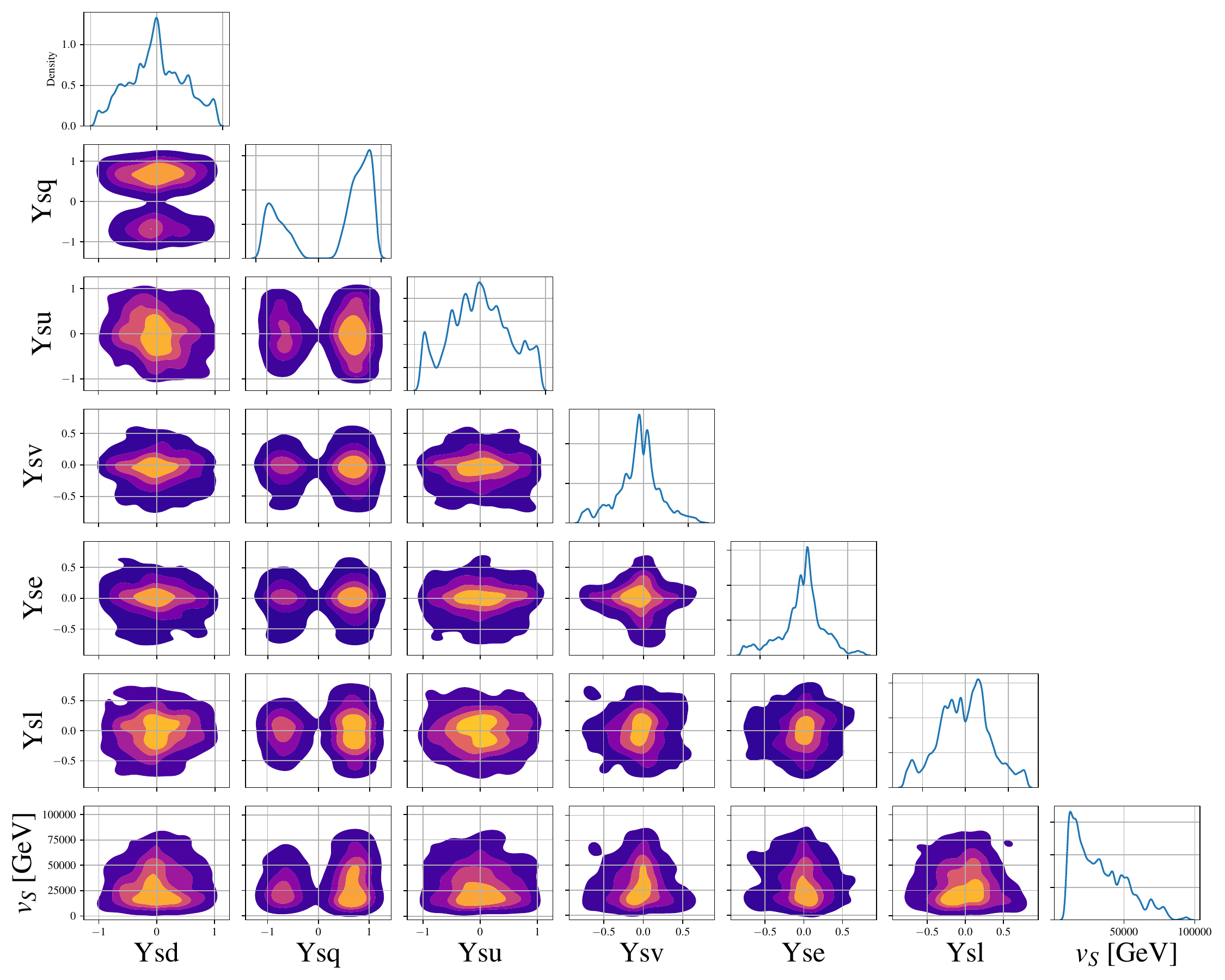}
    \caption{Correlations of Yukawa couplings in the MCMC scan.}
    \label{FIG:parcorner1}
\end{figure*}

\begin{figure*}[p]
    \centering
    \includegraphics[width=0.72\textwidth]{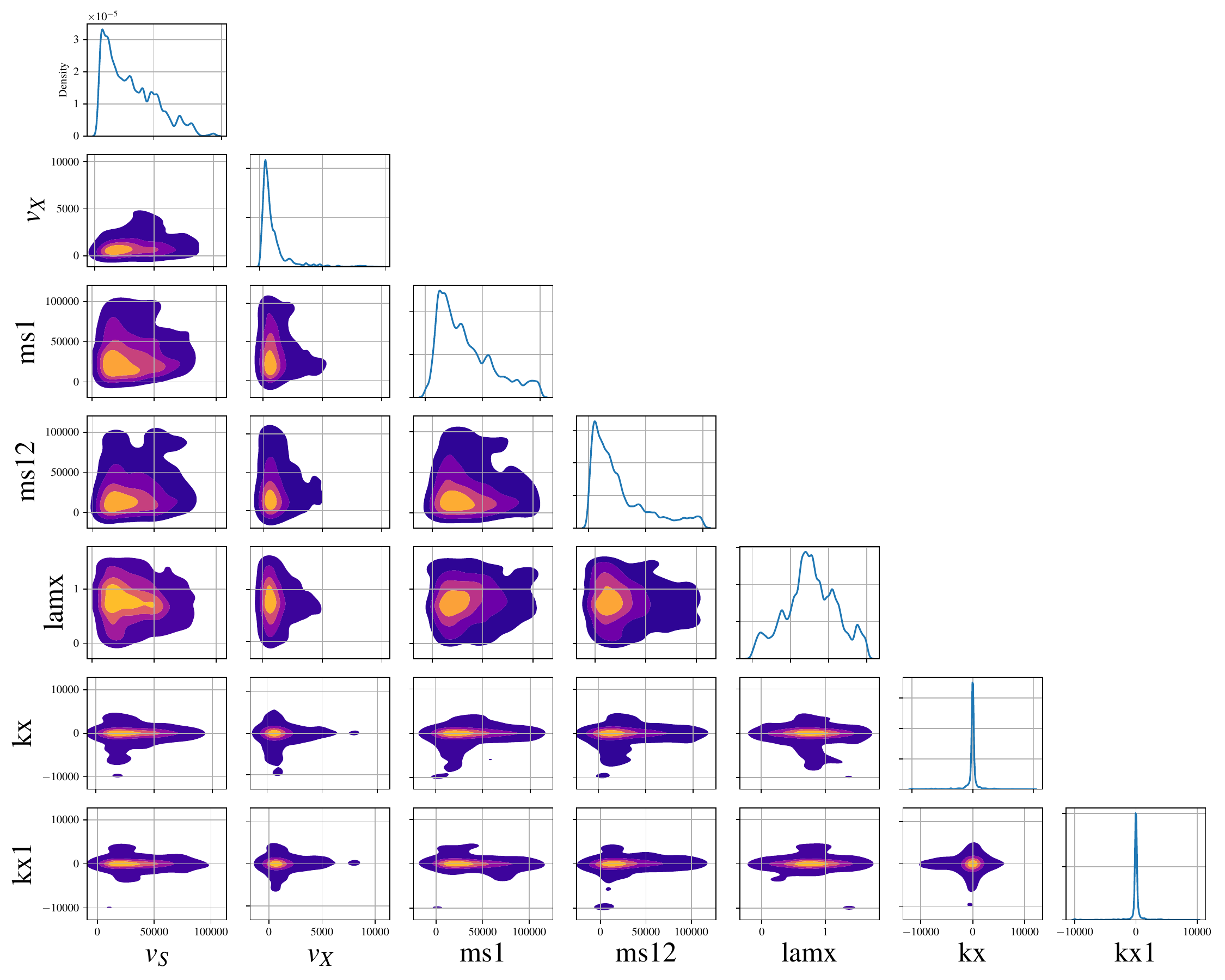}
    \caption{Correlations of dimensionful parameters in the MCMC scan.}
    \label{FIG:parcorner2}
\end{figure*}

\begin{figure*}[p]
    \centering
    \includegraphics[width=0.72\textwidth]{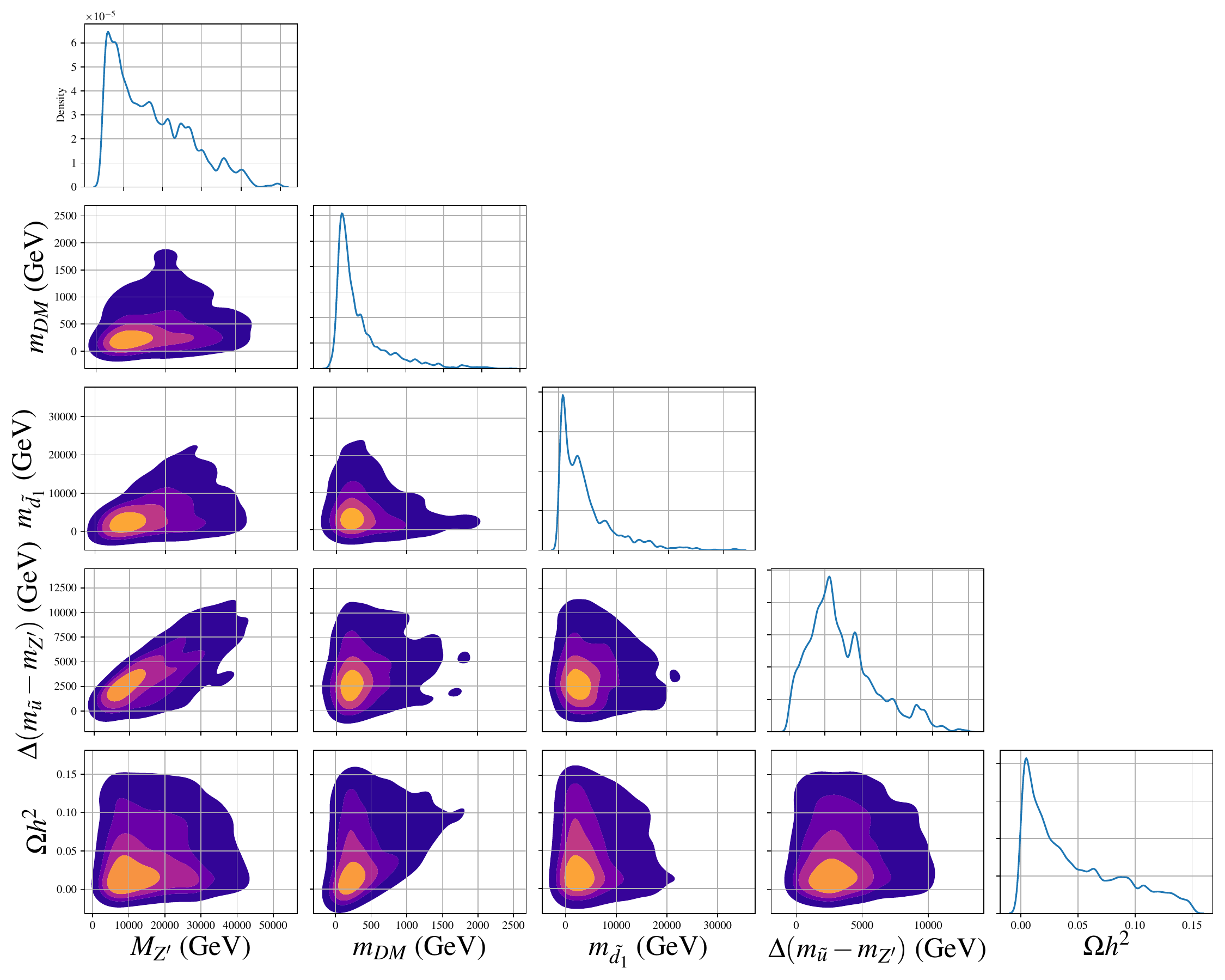}
    \caption{Correlations of selected observables in the MCMC scan.}
    \label{FIG:obscorner}
\end{figure*}

The scan identifies many points consistent with the target scenario: a heavy
baryonic \(Z^\prime\), with at least one secluded quark heavier than the
\(Z^\prime\), a dark matter relic density at or below the observed value, and
compatibility with direct-detection, Higgs, and collider constraints. The
distributions of the affine-invariant MCMC points in parameter space are shown
in Figs.~\ref{FIG:parcorner1} and~\ref{FIG:parcorner2}, while correlations
among selected observables are presented in
Fig.~\ref{FIG:obscorner}.

Among the observed correlations, the most significant is that between the dark
matter mass and the relic density. In the viable points, the dark matter
candidate is the lightest Majorana secluded neutrino,
\(\tilde{\nu}_1\). Its annihilation can proceed through
\(Z^\prime\)-mediated processes or through Yukawa interactions. Away from the
\(Z^\prime\) resonance, additional annihilation mechanisms become necessary to
reduce the relic density: some points rely on co-annihilation with secluded
charged leptons or down-type quarks, whereas others annihilate efficiently into
pairs of the light pseudoscalar \(A_1\).

To illustrate the viable regions, we select four representative benchmark
points. Their input parameters are listed in
Table~\ref{TAB:benchmarkinputs}, while their spectra and dark matter
observables are given in Table~\ref{TAB:benchmarkoutputs}. All four benchmarks
satisfy the selection criteria imposed in the scan.


\paragraph{Annihilation into \(A_1\) pairs.}

For a substantial part of the viable parameter space,
\(\tilde\nu_1\tilde\nu_1\to A_1A_1\) is the dominant annihilation
channel. The relic density and the individual channel contributions
quoted below are computed numerically with \MicrOMEGAs, using the
spectrum and vertices generated by \SARAH/\SPheno. The simplified
expressions below are intended only to illustrate the parametric
dependence and the \(p\)-wave suppression of the relevant amplitudes.

Two classes of diagrams contribute at leading order: \(t/u\)-channel
exchange of neutral secluded fermions and \(s\)-channel exchange of a
scalar, in particular the mostly-\(X_1\) state \(h_2\). In the
simplified limits considered below, both contributions are
\(p\)-wave suppressed, which also suppresses the corresponding
indirect-detection signal.

For a single Majorana fermion \(\chi\) coupled diagonally to a real
pseudoscalar \(A_1\) through an effective pseudoscalar coupling
\(g_{\rm eff}\), the leading \(p\)-wave thermally averaged
\(t/u\)-channel cross section is
\begin{equation}
    \left\langle \sigma v \right\rangle_{A_1A_1}
    \simeq
    \frac{g_{\rm eff}^4}{24\pi}
    \frac{m_\chi\left(m_\chi^2-m_{A_1}^2\right)^{5/2}}
    {\left(2m_\chi^2-m_{A_1}^2\right)^4}
    \frac{6}{x},
    \label{eq:AA_tu_estimate}
\end{equation}
which agrees with the corresponding simplified expression of
Ref.~\cite{Dolan:2014ska} in the Dirac case.

Neglecting the pseudoscalar mass, this gives
\begin{align}
  \Omega h^2 \approx
  1.4 \times 10^{-6}\,
  g_{\rm eff}^{-4}
  \left(\frac{m_{\tilde\nu_1}}{\mathrm{GeV}}\right)^2 .
\end{align}


For our model, the effective coupling is not straightforward to extract here since the physical state \(\tilde\nu_1\) is a mixture of several neutral fermions: while the coupling is involves $Y_{Xee}$ and $Y_{X\nu\nu}$, we have to diagonalise from the flavour to the mass basis (and of course this is only a simplifation of the \MicrOMEGAs result, which takes many more processes/vertices into account). But in general we expect $g_{\rm eff}$ to be less than one, and for this to dominate we need light $m_{\tilde{\nu}_1}$.

For the \(s\)-channel scalar contribution we model the effective
interactions as
\begin{align}
\mathcal{L}
\supset
-
\frac{y}{2}\,
\overline{\tilde{\nu}_1}\tilde{\nu}_1 h_2
-
\frac{\kappa}{2}\,
h_2A_1A_1 .
\end{align}
This gives
\begin{align}
\langle\sigma v\rangle
\simeq
\frac{1}{x}
\frac{3y^2\kappa^2}{64\pi}
\frac{\sqrt{1-m_{A_1}^2/m_{\tilde\nu_1}^2}}
     {(4m_{\tilde\nu_1}^2-m_{h_2}^2)^2},
\end{align}
which, neglecting \(m_{A_1}\), yields
\begin{align}
\Omega h^2
\approx
4.6\times10^{-7}
\left(
\frac{m_{h_2}^2-4m_{\tilde\nu_1}^2}
     {\mathrm{GeV}^2}
\right)^2
\left(
\frac{\mathrm{GeV}}
     {y\kappa}
\right)^2 .
\end{align}
Here \(\kappa\simeq\lambda_Xv_X\), which is typically of the same
order as \(m_{h_2}\).

In Benchmarks~2 and~4,
\(\tilde\nu_1\tilde\nu_1\to A_1A_1\) is the dominant annihilation
channel. In both cases \(m_{h_2}\) lies close to
\(2m_{\tilde\nu_1}\), suggesting that the enhancement is dominated by
resonant \(s\)-channel exchange of the mostly-\(X_1\) scalar, rather
than by the simplified \(t/u\)-channel contribution of
Eq.~\eqref{eq:AA_tu_estimate}. The relative contributions quoted for
the benchmark points are obtained from the full
\MicrOMEGAs calculation.

\paragraph{Co-annihilation with the secluded down-type quark.}

In Benchmark~3 the lightest secluded quark \(\tilde d_1\) lies only
\(40\,\mathrm{GeV}\) above the dark matter candidate,
\(\delta m/m_{\tilde\nu_1}\simeq3.3\%\). The corresponding Boltzmann
suppression factor at freeze-out is
\(r=\exp(-x_f\,\delta m/m_{\tilde\nu_1})\simeq0.44\), with
\(x_f\simeq25\), so the \(\tilde d_1\) population remains sufficiently
abundant during freeze-out to contribute significantly to the effective
annihilation rate. The colour-triplet \(\tilde d_1\) annihilates
predominantly into gluons through its QCD coupling, with
\[
\langle\sigma v\rangle_{\tilde d_1\tilde d_1^*\to gg}
\sim
\frac{14\pi\alpha_s^2}{9\,m_{\tilde d_1}^2}
\simeq
3\times10^{-25}\ {\rm cm}^3/{\rm s}.
\]
Weighted by the Griest--Seckel factor
\(r^2/(1+r)^2\simeq0.09\), this gives an effective cross section
\(\langle\sigma v\rangle_{\rm eff}\sim3\times10^{-26}\ {\rm cm}^3/{\rm s}\),
consistent with the value required to reproduce
\(\Omega h^2\simeq0.12\).

\paragraph{Annihilation into gauge bosons.}

In Benchmark~1 the dominant annihilation channel is
\(\tilde\nu_1\tilde\nu_1\to W^+W^-\), which is kinematically accessible
because the dark matter mass,
\(208\,\mathrm{GeV}\), lies above threshold. Since the dark matter
particle consists entirely of electroweak singlet states, there is no
direct coupling to the \(W\) bosons. The process therefore proceeds
through \(s\)-channel exchange of \(h_2\), and the benchmark lies close
to the resonance,
\(m_{h_2}\approx2m_{\tilde\nu_1}\).

The pseudoscalar \(A_1\) is light compared with the characteristic mass
scale of the secluded sector. Its mixing with the SM-like Higgs sector
is negligible, and its pseudoscalar nature suppresses its contributions
to the invisible widths of both the Higgs boson and the \(Z\).

The benchmark points are also compatible with current dark matter
direct-detection bounds. Indirect-detection constraints are expected to
be weak for the benchmark points presented here, either because the
dominant annihilation channel is \(p\)-wave suppressed or because the
dark matter is sufficiently heavy. A dedicated indirect-detection
analysis is left for future work, once the corresponding functionality
has been implemented in \BSMArt.

\begin{figure*}[p]
  \centering
  \includegraphics[width=0.75\textwidth]{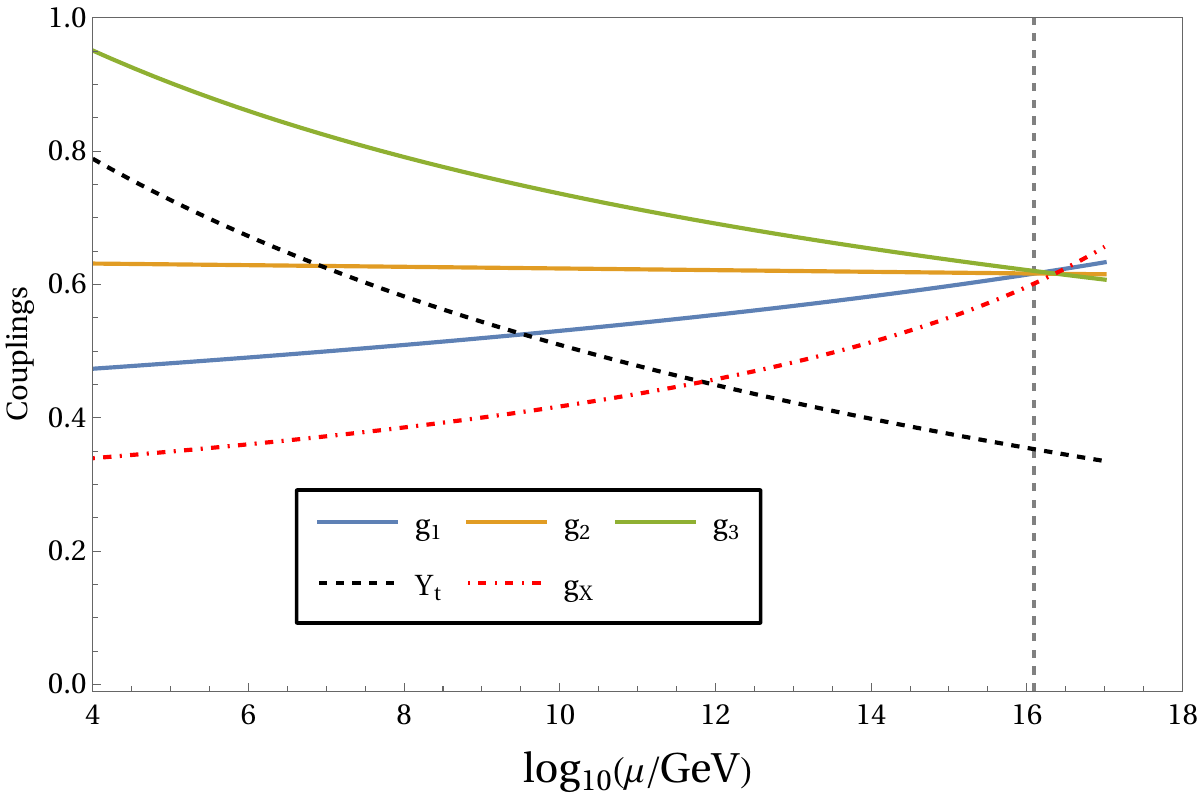}
  \caption{\label{FIG:gaugerunning352}Running of gauge couplings (plus the top Yukawa coupling $Y_t$ for comparison) in a
  representative viable point.}
\end{figure*}

Figure~\ref{FIG:gaugerunning352} shows the running of the gauge
couplings, the top Yukawa coupling, and the Higgs quartic for a typical
viable point. The dimensionless couplings are extracted from the
\SPheno output at the BSM scale,
\(10^{4}\,\mathrm{GeV}\), and subsequently evolved in
{\tt Mathematica} using the two-loop RGEs generated by \SARAH. The
gauge couplings evolve towards a common value at high scales.

\subsection{Non-unified scenario}
\label{sec:nonunified}

The benchmark scenarios discussed above were obtained under the simplifying
assumption that the \(U(1)_{\tilde B}\) gauge coupling unifies with the
Standard Model gauge couplings at the high scale. This provides a convenient
reference point, but it is not a necessary consequence of an ultraviolet
completion.

Many ultraviolet completions, including string compactifications, allow the
three Standard Model gauge couplings to unify while an additional Abelian
factor need not share the same high-scale coupling. An extra hidden
\(U(1)\), originating from a different sector of the underlying theory, may
therefore remain independent below the compactification scale. Keeping the
successful unification of the Standard Model gauge couplings intact, we
therefore consider the possibility that \(g_X\) is an independent parameter
at the high scale.

This freedom opens up a qualitatively different collider phenomenology. In
particular, it naturally allows a much larger hierarchy between the baryonic
\(Z^\prime\) and the secluded spectrum. Since the baryonic \(Z^\prime\) mass
is controlled by the product \(g_Xv_S\), whereas the secluded fermion masses
are governed primarily by Yukawa couplings times the same symmetry-breaking
scale, reducing \(g_X\) makes it possible to keep the \(Z^\prime\) at the TeV
scale while pushing part of the secluded spectrum to much higher masses.
Among the secluded states, the quarks are the easiest to move into the
multi-TeV regime while maintaining viable phenomenology. Such a spectrum
would be uncovered only in stages: the baryonic \(Z^\prime\) could be
discovered first, while the remaining anomaly-cancelling states would become
accessible only at higher collider energies.

To illustrate this possibility, we perform a second scan in which \(g_X\) is
treated as an independent input parameter at the high scale. The scan
follows the same strategy as before, with a linear prior for \(g_X\) and an
additional bias on the mass ratio \(m_{\tilde u}/m_{Z^\prime}\), in order to
favour spectra with secluded quarks substantially heavier than the baryonic
\(Z^\prime\).

For this scan we include \HiggsTools and {\sc ZPEED}~\cite{Kahlhoefer:2019vhz}
in the toolchain, together with the {\sc ZprimeExplorer}
data~\cite{Alvarez:2020yim,Lozano:2021zbu} implemented in \BSMArt.
Reducing \(g_X\) weakens the collider limits on the baryonic \(Z^\prime\),
allowing viable points with \(Z^\prime\) masses below the limits usually
quoted for sequential \(Z^\prime\) bosons or for models with comparable
gauge couplings.

The resulting non-unified benchmarks are listed in
Tables~\ref{TAB:NUInputs} and~\ref{TAB:NUOutputs}. They realise the target
scenario in which the secluded quarks lie in the multi-TeV to
tens-of-TeV range, while the baryonic \(Z^\prime\) remains at the TeV
scale.

The benchmarks also illustrate that viable points exist in which
\(m_{A_1}>m_{\rm DM}\), so that dark matter annihilation proceeds
predominantly into \(W^+W^-\). Nevertheless, annihilation into
\(A_1A_1\) remains the more typical scenario across the viable parameter
space, and in this case the dark matter candidate can naturally be much
heavier.

\begin{table*}[tbp]
\centering
\renewcommand{\arraystretch}{1.15}
\setlength{\tabcolsep}{10pt}
\begin{tabular}{lcccc}
\toprule
Parameter & Benchmark 1 & Benchmark 2 & Benchmark 3 & Benchmark 4 \\
\midrule
\(Y_{sd}\)   & 0.0299 & 0.5517 & 0.0153 & -0.2988 \\
\(Y_{sq}\)   & 0.7286 & 0.8671 & -0.5342 & -0.8130 \\
\(Y_{su}\)   & -0.8990 & 0.1203 & 0.2414 & -0.2238 \\
\(Y_{s\nu}\)   & -0.6184 & 0.1532 & 0.0513 & -0.0732 \\
\(Y_{se}\)   & 0.0205 & -0.4177 & -0.2165 & 0.3881 \\
\(Y_{sl}\)   & 0.6070 & -0.1197 & -0.1156 & 0.3001 \\
\(\lambda_{HS}\) & 0.2834 & 0.5572 & 0.1448 & 0.1554 \\
\(\lambda_{S}\)  & 0.2881 & 0.9681 & 0.4317 & 1.4516 \\
\(v_S\)   & $6.878 \times 10^4$ & $4.317 \times 10^4$ &
$7.184 \times 10^4$ & $2.644 \times 10^4$ \\
\(v_X\)   & 555.8536 & 3248.5315 & 457.9725 & 856.9767 \\
\(y_{off}\) & -0.3874 & 0.4113 & 0.2160 & -0.0974 \\
\(m_{S1}\)  & $1.845 \times 10^4$ & $1.726 \times 10^4$ &
$5.404 \times 10^4$ & $6.494 \times 10^4$ \\
\(m_{S12}\)  & $8.926 \times 10^4$ & $6.252 \times 10^4$ &
$2.431 \times 10^4$ & 8537.6023 \\
\(Y_{Xee}\)  & 0.6884 & -0.5005 & -0.6351 & -0.7042 \\
\(\lambda_{X}\)  & 0.3894 & 0.5225 & 0.6782 & 0.6866 \\
\(k_X\)    & 64.5405 & -354.5002 & 118.0836 & 19.1421 \\
\(k_{X1}\)   & 259.7518 & -1231.0002 & -208.4170 & -10.8663 \\
\bottomrule
\end{tabular}
\caption{\label{TAB:benchmarkinputs}
Input parameters for the four benchmark points.}
\end{table*}

\begin{table*}[tbp]
\centering
\renewcommand{\arraystretch}{1.15}
\setlength{\tabcolsep}{8pt}
\begin{tabular}{lcccc}
\toprule
 & Benchmark 1 & Benchmark 2 & Benchmark 3 & Benchmark 4 \\
\midrule
\multicolumn{5}{l}{\textit{Masses (GeV)}} \\
\midrule
$A_1$ & $2.21 \times 10^{2}$ & $1.76 \times 10^{2}$ & $1.73 \times 10^{2}$ &
$9.17 \times 10^{1}$ \\
$h_2$ & $4.26 \times 10^{2}$ & $2.15 \times 10^{3}$ & $3.45 \times 10^{2}$ &
$6.36 \times 10^{2}$ \\
$h_3$ & $4.48 \times 10^{4}$ & $3.25 \times 10^{4}$ & $4.93 \times 10^{4}$ &
$2.12 \times 10^{4}$ \\
$\tilde{d}_1$ & $1.55 \times 10^{3}$ & $3.79 \times 10^{3}$ &
$1.26 \times 10^{3}$ & $4.59 \times 10^{3}$ \\
$\tilde{d}_2$ & $3.57 \times 10^{4}$ & $1.57 \times 10^{4}$ &
$1.73 \times 10^{4}$ & $6.03 \times 10^{3}$ \\
$\tilde{d}_3$ & $3.84 \times 10^{4}$ & $2.86 \times 10^{4}$ &
$4.46 \times 10^{4}$ & $1.85 \times 10^{4}$ \\
$\tilde{u}$ & $-3.84 \times 10^{4}$ & $-2.86 \times 10^{4}$ &
$4.46 \times 10^{4}$ & $1.85 \times 10^{4}$ \\
$\tilde{e}$ & $1.57 \times 10^{4}$ & $-2.11 \times 10^{3}$ &
$-5.03 \times 10^{3}$ & $3.50 \times 10^{3}$ \\
$\tilde{\nu}_1$ & $-2.08 \times 10^{2}$ & $1.03 \times 10^{3}$ &
$1.22 \times 10^{3}$ & $3.06 \times 10^{2}$ \\
$\tilde{\nu}_2$ & $5.57 \times 10^{2}$ & $-2.79 \times 10^{3}$ &
$-1.50 \times 10^{3}$ & $-8.59 \times 10^{2}$ \\
$\tilde{\nu}_3$ & $-9.78 \times 10^{3}$ & $3.79 \times 10^{3}$ &
$5.56 \times 10^{3}$ & $2.43 \times 10^{3}$ \\
$\tilde{\nu}_4$ & $1.01 \times 10^{4}$ & $-5.54 \times 10^{3}$ &
$-5.84 \times 10^{3}$ & $-2.98 \times 10^{3}$ \\
$\tilde{\nu}_D$ & $-1.57 \times 10^{4}$ & $2.11 \times 10^{3}$ &
$5.03 \times 10^{3}$ & $-3.50 \times 10^{3}$ \\
$Z^\prime$ & $3.54 \times 10^{4}$ & $2.24 \times 10^{4}$ &
$3.69 \times 10^{4}$ & $1.35 \times 10^{4}$ \\
$\phi_1$ & $1.26 \times 10^{4}$ & $8.95 \times 10^{3}$ &
$3.18 \times 10^{4}$ & $6.65 \times 10^{3}$ \\
$\phi_2$ & $4.60 \times 10^{4}$ & $3.12 \times 10^{4}$ &
$4.82 \times 10^{4}$ & $6.62 \times 10^{4}$ \\
$\sigma_1$ & $1.26 \times 10^{4}$ & $8.93 \times 10^{3}$ &
$3.18 \times 10^{4}$ & $6.65 \times 10^{3}$ \\
$\sigma_2$ & $4.60 \times 10^{4}$ & $3.12 \times 10^{4}$ &
$4.82 \times 10^{4}$ & $6.62 \times 10^{4}$ \\
\midrule
\multicolumn{5}{l}{\textit{Dark matter observables}} \\
\midrule
$\Omega h^2$ & $1.19 \times 10^{-1}$ & $1.04 \times 10^{-1}$ &
$1.16 \times 10^{-1}$ & $6.84 \times 10^{-3}$ \\
DM mass (GeV) & $2.08 \times 10^{2}$ & $1.03 \times 10^{3}$ &
$1.22 \times 10^{3}$ & $3.06 \times 10^{2}$ \\
DM ann.\ channel & $W^\pm W^\mp$ & $A_1 A_1$ &
Coann.\ with $\tilde{d}$ & $A_1 A_1$ \\
\bottomrule
\end{tabular}
\caption{\label{TAB:benchmarkoutputs}
Spectrum and dark matter observables for the four benchmark points.
Negative fermion masses follow the \SPheno convention and correspond to signs
of the mass eigenvalues; physical masses are their absolute values.}
\end{table*}

\begin{table*}[tbp]
\centering
\renewcommand{\arraystretch}{1.15}
\setlength{\tabcolsep}{10pt}
\begin{tabular}{lccc}
\toprule
Coupling & NU Benchmark 1 & NU Benchmark 2 & NU Benchmark 3 \\
\midrule
\(Y_{sd}\) & 0.859108 & -0.833114 & -0.276972 \\
\(Y_{sq}\) & -0.144561 & 0.350448 & -0.653804 \\
\(Y_{su}\)& -0.378541 & -0.101361 & 0.644732 \\
\(Y_{s\nu}\) & 0.032603 & 0.463943 & -0.005107 \\
\(Y_{se}\) & -0.296013 & -0.047203 & -0.624201 \\
\(Y_{sl}\) & 0.117179 & 0.248818 & 0.027715 \\
\(\lambda_{HS}\) & 0.391839 & 0.319604 & 0.200522 \\
\(\lambda_S\) & 0.575919 & 0.801504 & 0.068342 \\
\(v_S\) & 69398.230100 & 35556.607200 & 89144.592400 \\
\(v_X\) & 1965.017940 & 1283.169090 & 24.451389 \\
\(y_{off}\) & 0.034726 & 0.156220 & 0.209267 \\
\(m_{S1}\) & 26185.365200 & 9119.252370 & 30180.806300 \\
\(m_{S12}\) & 30183.062200 & 7874.798990 & 1637.788200 \\
\(Y_{Xee}\) & -0.261437 & 0.513247 & 0.790335 \\
\(\lambda_X\) & 1.180236 & 0.079999 & 0.747554 \\
\(k_X\) & -55.991561 & -13.442808 & -753.167520 \\
\(k_{X1}\) & 46.574828 & -12.640258 & -498.417466 \\
\(\lambda_H\) & -0.013322 & 0.009375 & -0.017711 \\
$g_X$ & 0.058370 & 0.116726 & 0.027463 \\
\bottomrule
\end{tabular}
\caption{\label{TAB:NUInputs}
Input parameters for the three non-unified (NU) benchmark points.}
\end{table*}

\begin{table*}[tbp]
\centering
\renewcommand{\arraystretch}{1.15}
\setlength{\tabcolsep}{9pt}
\begin{tabular}{lccc}
\toprule
 & NU Benchmark 1 & NU Benchmark 2 & NU Benchmark 3 \\
\midrule
\multicolumn{4}{l}{\textit{Masses (GeV)}} \\
\midrule
$A_1$ & $1.45 \times 10^{2}$ & $6.84 \times 10^{1}$ & $3.05 \times 10^{2}$ \\
$h_2$ & $1.26 \times 10^{3}$ & $6.94 \times 10^{2}$ & $2.33 \times 10^{2}$ \\
$h_3$ & $4.76 \times 10^{4}$ & $2.31 \times 10^{4}$ & $5.42 \times 10^{4}$ \\
$\tilde{d}_1$ & $1.30 \times 10^{4}$ & $3.47 \times 10^{3}$ & $1.78 \times 10^{4}$ \\
$\tilde{d}_2$ & $2.37 \times 10^{4}$ & $1.46 \times 10^{4}$ & $3.77 \times 10^{4}$ \\
$\tilde{d}_3$ & $4.69 \times 10^{4}$ & $2.30 \times 10^{4}$ & $4.87 \times 10^{4}$ \\
$\tilde{u}$ & $1.30 \times 10^{4}$ & $-1.46 \times 10^{4}$ & $4.87 \times 10^{4}$ \\
$\tilde{e}$ & $4.70 \times 10^{3}$ & $4.71 \times 10^{3}$ & $1.12 \times 10^{3}$ \\
$\tilde{\nu}_1$ & $6.00 \times 10^{2}$ & $-3.16 \times 10^{2}$ & $-1.09 \times 10^{2}$ \\
$\tilde{\nu}_2$ & $-1.26 \times 10^{3}$ & $1.01 \times 10^{3}$ & $1.25 \times 10^{2}$ \\
$\tilde{\nu}_3$ & $7.52 \times 10^{3}$ & $-5.13 \times 10^{3}$ & $-1.44 \times 10^{4}$ \\
$\tilde{\nu}_4$ & $-8.18 \times 10^{3}$ & $5.81 \times 10^{3}$ & $1.44 \times 10^{4}$ \\
$\tilde{\nu}_D$ & $-4.70 \times 10^{3}$ & $-4.71 \times 10^{3}$ & $-1.12 \times 10^{3}$ \\
$Z^\prime$ & $6.21 \times 10^{3}$ & $6.08 \times 10^{3}$ & $3.85 \times 10^{3}$ \\
$\phi_1$ & $2.54 \times 10^{4}$ & $9.03 \times 10^{3}$ & $2.76 \times 10^{4}$ \\
$\phi_2$ & $2.89 \times 10^{4}$ & $1.09 \times 10^{4}$ & $3.26 \times 10^{4}$ \\
$\sigma_1$ & $2.54 \times 10^{4}$ & $9.02 \times 10^{3}$ & $2.76 \times 10^{4}$ \\
$\sigma_2$ & $2.89 \times 10^{4}$ & $1.09 \times 10^{4}$ & $3.26 \times 10^{4}$ \\
\midrule
\multicolumn{4}{l}{\textit{Dark matter observables}} \\
\midrule
$\Omega h^2$ & $7.53 \times 10^{-2}$ & $1.04 \times 10^{-1}$ & $2.89 \times 10^{-2}$ \\
DM Mass (GeV) & $6.00 \times 10^{2}$ & $3.16 \times 10^{2}$ & $1.09 \times 10^{2}$ \\
DM ann.\ channel & $A_1, A_1$ & $A_1, A_1$   & $W^\pm W^\mp$ \\
\bottomrule
\end{tabular}
\caption{\label{TAB:NUOutputs}
Spectrum and dark matter observables for the three non-unified (NU) benchmark
points. Negative fermion masses follow the \SPheno convention and correspond to
signs of the mass eigenvalues; physical masses are their absolute values.}
\end{table*}

\clearpage

\section{Conclusions}
\label{sec:conclusion}

We have constructed and studied an anomaly-free baryonic completion of the
Standard Model. The new Abelian gauge symmetry acts as ordinary baryon number
on the Standard Model fields and is extended to a secluded sector whose charges
are fixed by anomaly cancellation and by the allowed renormalisable
interactions. The corresponding massive gauge boson is therefore naturally
interpreted as a baryonic \(Z^\prime\), but the organising principle is the
anomaly-free completion of \(U(1)_B\), rather than the introduction of a
generic extra neutral gauge boson.

Anomaly cancellation alone does not define a viable model. For the class of
completions considered here, the minimal construction with secluded fermions
and a single scalar responsible for \(U(1)_{\tilde B}\) breaking is not
phenomenologically viable. It leaves accidental global symmetries that render
some exotic states stable, including coloured states. Removing these unwanted
relics requires additional singlet scalars and Yukawa couplings. The enlarged
scalar sector is therefore essential to the consistency of the model: it
allows the exotic states to decay, preserves a stable neutral
\(Z_2^{\rm DM}\)-odd state, and determines the structure of the neutral
secluded sector.

We focus here on one representative charge assignment from the family of
anomaly-free solutions described in
appendix~\ref{app:anomaly_constraints}. Each charge assignment in this family
defines a distinct model requiring a dedicated implementation and
phenomenological analysis. The model was implemented in \SARAH, with a
\SPheno spectrum generator and phenomenological analysis using
\MicrOMEGAs, \SModelS, and \HiggsTools.

The allowed parameter space is substantially more restricted than suggested by
the anomaly equations alone. The requirements of a sufficiently heavy
\(Z^\prime\), perturbative couplings up to high scales, acceptable
exotic-particle decays, compatibility with collider and Higgs constraints, and
the observed dark-matter abundance impose strong correlations among the
secluded Yukawa couplings, the scalar spectrum, and the new gauge coupling.
The regime in which part of the secluded spectrum is heavier than the
\(Z^\prime\) is realised only in restricted regions of parameter space,
requiring some secluded Yukawa couplings to be large enough to raise the
masses of the heavy exotic states while others remain small enough to preserve
a viable neutral dark matter candidate and perturbativity.

The dark-matter phenomenology is not that of a minimal
\(Z^\prime\)-portal model. Although annihilation through the baryonic
\(Z^\prime\) contributes to the relic density, the viable regions are also
controlled by co-annihilations with nearby secluded states or by annihilation
into the light pseudoscalar \(A_1\). The scalar fields introduced to ensure
the consistency of the model therefore also play a direct role in the
dark-matter dynamics.

The same field content improves the high-scale evolution of the Standard Model
gauge couplings. We have not assumed an embedding into a simple Grand Unified
Theory, but instead required compatibility with a common high-scale gauge
coupling, as may arise in ultraviolet completions where gauge couplings are
controlled by common moduli. Within this framework, the secluded sector
simultaneously cancels the baryonic anomalies and improves the gauge-coupling
evolution relative to the Standard Model.

A precise spectrum calculation proved essential. The collective breaking of
the accidental symmetries is sufficiently intricate that the pseudoscalar mass
cannot be treated as an independent phenomenological parameter: the two-loop
contribution to the pole mass is numerically significant. This underlines the
importance of precise spectrum generators for models with extended scalar
sectors and non-trivial symmetry-breaking patterns.

Several questions remain open. The neutrino sector deserves a more complete
analysis than the leading treatment adopted here. The full model contains
additional neutral fermions and singlet interactions, and a Majorana mass for
\(\nu_R\) is allowed by the gauge symmetries. These ingredients make
seesaw-like neutrino masses natural, but the complete loop-corrected
neutral-fermion mass matrix has not been studied in this work.

Proton stability deserves a dedicated analysis at the level of
higher-dimensional operators. The extended baryonic gauge symmetry forbids the
usual proton-decay operators built solely from Standard Model fields before
\(U(1)_{\tilde B}\) breaking. After symmetry breaking, however, singlet
insertions may generate effective baryon-number-violating operators; we have
assumed that these are suppressed by the ultraviolet scale, by small singlet
expectation values, or by additional selection rules.

The model also has potentially rich cosmological implications. The heavy
secluded states, together with the quark and lepton portal interactions, could
participate in asymmetry transfer or baryogenesis, while the same interactions
could wash out a pre-existing asymmetry if they remain in equilibrium at the
relevant temperatures. The singlet scalar sector may also lead to a
non-trivial finite-temperature history, including multi-step or first-order
phase transitions accompanied by a stochastic gravitational-wave background.
The study of baryogenesis, washout constraints, and possible
gravitational-wave signals is left for future work.

The present work provides an explicit benchmark for a non-supersymmetric
anomaly-free completion of gauged baryon number. In this benchmark the
secluded sector simultaneously cancels the anomalies, contains the dark matter
candidate, and modifies the high-scale evolution of the gauge couplings; the
same interactions required to avoid stable exotic relics also determine the
neutral sector and the dominant dark-matter annihilation mechanisms. In the
region of parameter space studied here, the baryonic \(Z^\prime\) may be
lighter than part of the secluded spectrum, so that the first experimental
evidence for the new gauge boson would not immediately reveal the full
anomaly-free completion. This provides a well-defined target for future
high-energy hadron colliders and a starting point for systematic studies of
the wider family of anomaly-free baryonic completions.



\appendix

\section{Anomaly constraints and charge assignments}
\label{app:anomaly_constraints}

In this appendix we present the anomaly analysis underlying the charge
assignments used in the main text. We show how the anomaly-free baryonic
completion emerges from a restricted but non-trivial class of solutions, and
provide an additional charge assignment to illustrate that the benchmark
studied in the main text is only one representative of a broader family.

The new gauge factor is denoted by \(\UB\). Its corresponding charge,
\(\tB\), coincides with ordinary baryon number on the Standard Model fields in
the benchmark model. Throughout the anomaly analysis below, however, we keep
the more general notation
\[
    z_Q,\ z_u,\ z_d,\ z_L,\ z_e,\ z_\nu,\ z_H ,
\]
for the \(\tB\) charges of the SM fields, in order to display the structure of
the anomaly-free solutions before specialising to the baryonic branch.

Throughout this appendix we work in a left-chiral Weyl basis and make the
following assumptions:
\begin{itemize}
    \item SM fermion masses arise from the usual Yukawa couplings to the Higgs
    doublet \(H\), including Dirac neutrino Yukawa couplings.
    \item The SM \(\tB\) charges are family universal.
    \item The secluded sector is vector-like with respect to
    \(SU(3)_c\times SU(2)_L\times U(1)_Y\), but chiral under \(\UB\).
    \item Secluded-sector fermion masses arise from Yukawa couplings to a
    singlet scalar \(S\), whose vacuum expectation value breaks \(\UB\).
    \item Only renormalisable Yukawa interactions are considered.
\end{itemize}

The field content used in the anomaly analysis is summarised in
Table~\ref{tab:app_particle_content}. The index \(f=1,2,3\) labels the SM
families, while \(i,j,k,l\) denote possible secluded-sector multiplicities.

\begin{table*}[t]
\centering
\renewcommand{\arraystretch}{1.12}
\setlength{\tabcolsep}{10pt}
\begin{tabular}{lcccc}
\toprule
Field & \(SU(3)_c\) & \(SU(2)_L\) & \(U(1)_Y\) & \(\UB\) \\
\midrule
\multicolumn{5}{l}{\textit{Standard Model sector}} \\
\midrule
\(Q_L^f\)        & \(3\)      & \(2\) & \(1/6\)  & \(z_Q\) \\
\(u_R^{c,f}\)    & \(\bar 3\) & \(1\) & \(-2/3\) & \(z_u\) \\
\(d_R^{c,f}\)    & \(\bar 3\) & \(1\) & \(1/3\)  & \(z_d\) \\
\(L_L^f\)        & \(1\)      & \(2\) & \(-1/2\) & \(z_L\) \\
\(e_R^{c,f}\)    & \(1\)      & \(1\) & \(1\)    & \(z_e\) \\
\(\nu_R^{c,f}\)  & \(1\)      & \(1\) & \(0\)    & \(z_\nu\) \\
\(H\)            & \(1\)      & \(2\) & \(1/2\)  & \(z_H\) \\
\midrule
\multicolumn{5}{l}{\textit{Secluded sector}} \\
\midrule
\(\psi^L_{L,i}\)             & \(1\)      & \(2\) & \(y_L^i\)   & \(q_L^i\) \\
\((\psi^L_{R,i})^c\)         & \(1\)      & \(2\) & \(-y_L^i\)  & \(\tilde q_L^i\) \\
\(\psi^l_{L,j}\)             & \(1\)      & \(1\) & \(y_l^j\)   & \(q_l^j\) \\
\((\psi^l_{R,j})^c\)         & \(1\)      & \(1\) & \(-y_l^j\)  & \(\tilde q_l^j\) \\
\(\psi^Q_{L,l}\)             & \(3\)      & \(2\) & \(y_Q^l\)   & \(q_Q^l\) \\
\((\psi^Q_{R,l})^c\)         & \(\bar 3\) & \(2\) & \(-y_Q^l\)  & \(\tilde q_Q^l\) \\
\(\psi^D_{L,k}\)             & \(3\)      & \(1\) & \(y_D^k\)   & \(q_D^k\) \\
\((\psi^D_{R,k})^c\)         & \(\bar 3\) & \(1\) & \(-y_D^k\)  & \(\tilde q_D^k\) \\
\(S\)                        & \(1\)      & \(1\) & \(0\)       & \(q_S\) \\
\bottomrule
\end{tabular}
\caption{\label{tab:app_particle_content}
Field content used in the anomaly analysis. The secluded sector is vector-like
under the SM gauge group but chiral under \(\UB\).}
\end{table*}

\subsection{Gauge-invariance constraints}
\label{app:gauge_invariance_constraints}

Gauge invariance of the SM Yukawa couplings
\[
    H Q_L u_R,\qquad
    \bar H Q_L d_R,\qquad
    \bar H L_L e_R,\qquad
    H L_L\nu_R
\]
imposes
\begin{equation}
\begin{aligned}
    z_e &= z_H-z_L, &
    z_\nu &= -z_H-z_L, \\
    z_d &= z_H-z_Q, &
    z_u &= -z_H-z_Q .
\end{aligned}
\label{eq:app_sm_yukawa_constraint}
\end{equation}

For the baryonic branch studied in the main text,
\[
    z_Q=\frac13,\qquad
    z_u=z_d=-\frac13,\qquad
    z_L=z_e=z_\nu=z_H=0,
\]
in the left-chiral Weyl convention, so that
\(\tB|_{\rm SM}=B\).

The secluded fermions acquire masses through Yukawa couplings to the singlet
scalar \(S\). At the renormalisable level, the coupling may involve either
\(S\) or \(S^\dagger\), so gauge invariance requires
\begin{equation}
    \tilde q_X = - q_X + \epsilon_X q_S,
    \qquad
    X\in\{L,l,Q,D\},
    \label{eq:app_sec_yukawa_constraint}
\end{equation}
where \(\epsilon_X=\pm1\). For \(q_S\neq0\), the two left-chiral fields
forming each Dirac pair do not carry opposite \(\UB\) charges. The secluded
fermions are therefore vector-like under the SM gauge group but chiral under
\(\UB\), and acquire masses only after \(S\) develops a vacuum expectation
value.

\subsection{General anomaly conditions}
\label{app:general_anomalies}

The anomalies involving only SM gauge factors cancel automatically, since the
SM sector is unchanged and the secluded sector is vector-like under the SM
gauge group. The non-trivial conditions are the mixed gauge, cubic, and
mixed gauge--gravitational anomalies involving \(\UB\). We define
\begin{align}
    \Tr[\tB]_{\rm SM}      &= -\Tr[\tB]_{\rm sec}      \equiv t_{\tilde B}, \\
    \Tr[Y^2 \tB]_{\rm SM}  &= -\Tr[Y^2 \tB]_{\rm sec}  \equiv t_{YY\tilde B}, \\
    \Tr[Y \tB^2]_{\rm SM}  &= -\Tr[Y \tB^2]_{\rm sec}  \equiv t_{Y\tilde B\tilde B}, \\
    \Tr[\tB^3]_{\rm SM}    &= -\Tr[\tB^3]_{\rm sec}    \equiv t_{\tilde B\tilde B\tilde B}, \\
    \Tr[\tB T_2^2]_{\rm SM}&= -\Tr[\tB T_2^2]_{\rm sec}\equiv t_2, \\
    \Tr[\tB T_3^2]_{\rm SM}&= -\Tr[\tB T_3^2]_{\rm sec}\equiv t_3,
\end{align}
where the traces run over all left-chiral fermions, and \(T_2\) and
\(T_3\) denote the generators of \(SU(2)_L\) and \(SU(3)_c\), respectively,
with the standard normalisation.

Before imposing the Yukawa relations, the SM contribution is
\begin{align}
    t_{\tilde B} &= 3\left(6z_Q+3z_u+3z_d+2z_L+z_e+z_\nu\right), \\
    t_{YY\tilde B} &= 3\left(\frac{z_Q}{6}+\frac{4z_u}{3}+\frac{z_d}{3}
    +\frac{z_L}{2}+z_e\right), \\
    t_{Y\tilde B\tilde B} &= 3\left(z_Q^2-2z_u^2+z_d^2-z_L^2+z_e^2\right), \\
    t_{\tilde B\tilde B\tilde B} &= 3\left(6z_Q^3+3z_u^3+3z_d^3
    +2z_L^3+z_e^3+z_\nu^3\right), \\
    t_2 &= 3\left(3z_Q+z_L\right), \\
    t_3 &= 3\left(2z_Q+z_u+z_d\right).
\end{align}

Using Eq.~\eqref{eq:app_sm_yukawa_constraint}, these expressions reduce to
\begin{equation}
\begin{aligned}
    t_{\tilde B} &= 0, &
    t_3 &= 0, \\
    t_{YY\tilde B} &= -\frac{1}{2}t_2, &
    t_{Y\tilde B\tilde B} &= -2z_H\,t_2, \\
    t_{\tilde B\tilde B\tilde B} &= -6z_H^2\,t_2 .
\end{aligned}
\label{eq:app_sm_anomaly_relations}
\end{equation}

After imposing the SM Yukawa constraints, the independent SM anomaly input
reduces to the mixed anomaly coefficient \(t_2\) together with the Higgs
charge \(z_H\). In the baryonic branch studied in the main text,
\(z_H=0\), so the anomaly matching simplifies considerably: the entire SM
contribution is encoded in the mixed \(SU(2)_L^2-\UB\) anomaly.

\subsection{SM-like organisation of the secluded sector}
\label{app:secluded_structure}

We now impose a restricted structure on the secluded sector, selecting a
tractable family of anomaly-free solutions with SM-like gauge
representations and suitable phenomenological properties, rather than
attempting a complete classification of anomaly-free baryonic
completions.

To cancel the mixed gravitational and \(SU(3)_c^2-\UB\) anomalies in the
same SM-like pattern, we impose
\begin{equation}
\begin{aligned}
    N_l &= 2N_L, &
    \epsilon_l &= -\epsilon_L, \\
    N_D &= 2N_Q, &
    \epsilon_D &= -\epsilon_Q .
\end{aligned}
\label{eq:app_NE_ND_conditions}
\end{equation}
We then split the secluded singlets into electron-like and neutrino-like
species, and the secluded colour triplets into two down-type species:
\begin{equation}
\begin{aligned}
    q_l^j &=
    \begin{cases}
        q_{E}, & j\le N_L,\\
        q_{\nu}, & j>N_L,
    \end{cases}
    &
    q_D^k &=
    \begin{cases}
        q_{d1}, & k\le N_Q,\\
        q_{d2}, & k>N_Q,
    \end{cases}
    \\
    q_Q^l &= q_Q,\qquad
    q_L^i = q_L,
\end{aligned}
\label{eq:app_charge_pattern}
\end{equation}
with analogous assignments for the hypercharges.

With this structure and the Yukawa relations of
Eq.~\eqref{eq:app_sec_yukawa_constraint}, the remaining secluded-sector
anomaly coefficients become
\begin{equation}
\begin{aligned}
-t_{YY\tilde B}={}&
q_S\Big[
  3\epsilon_Q N_Q
  \left(2y_Q^2-y_{d1}^2-y_{d2}^2\right)
\\
&\hspace{1.8cm}
  +\epsilon_L N_L
  \left(2y_L^2-y_E^2-y_\nu^2\right)
\Big],
\\[1mm]
-t_{Y\tilde B\tilde B}={}&
-q_S^2\Big[
  3N_Q\left(2y_Q+y_{d1}+y_{d2}\right)
\\
&\hspace{1.8cm}
  +N_L\left(2y_L+y_E+y_\nu\right)
\Big]
\\
&+2q_S\Big[
  3\epsilon_Q N_Q
  \left(2y_Q q_Q-y_{d1}q_{d1}-y_{d2}q_{d2}\right)
\\
&\hspace{1.8cm}
  +\epsilon_L N_L
  \left(2y_L q_L-y_Eq_E-y_\nu q_\nu\right)
\Big],
\\[1mm]
-t_{\tilde B\tilde B\tilde B}={}&
-3q_S^2\Big[
  3N_Q\left(2q_Q+q_{d1}+q_{d2}\right)
\\
&\hspace{1.8cm}
  +N_L\left(2q_L+q_E+q_\nu\right)
\Big]
\\
&+3q_S\Big[
  3\epsilon_Q N_Q
  \left(2q_Q^2-q_{d1}^2-q_{d2}^2\right)
\\
&\hspace{1.8cm}
  +\epsilon_L N_L
  \left(2q_L^2-q_E^2-q_\nu^2\right)
\Big].
\end{aligned}
\label{eq:app_sec_anomalies}
\end{equation}

The cubic and mixed anomalies contain quadratic charge combinations. To
avoid irrational square roots when solving the anomaly equations, we
restrict to
\begin{equation}
\begin{aligned}
    q_{d1} &= \pm(q_Q+z_H), &
    q_{d2} &= \pm(q_Q-z_H), \\
    q_E    &= \pm(q_L+z_H), &
    q_\nu  &= \pm(q_L-z_H).
\end{aligned}
\label{eq:app_rationality}
\end{equation}
This reduces the problem to a discrete set of sign choices together with
the continuous parameters
\[
    \{y_L,y_{E},y_{\nu},y_Q,y_{d1},y_{d2},
    z_L,z_Q,z_H,q_S,q_L,q_Q,\epsilon_L,\epsilon_Q\}.
\]

\subsection{Phenomenological filters}
\label{app:hypercharge_filter}

The anomaly equations admit many solutions. We narrow the allowed set by
imposing several phenomenological requirements.

Colour-singlet bound states should carry integer electric charge. This
favours hypercharges close to their SM values,
\[
    y_Q \sim \frac16,\quad
    y_{d1},y_{d2}\sim \frac13,\quad
    y_L\sim \frac12,\quad
    y_{E},y_{\nu}\sim 0 \ \text{or}\ 1.
\]
The additional matter should also improve the high-scale evolution of the
SM gauge couplings. As discussed in the main text, we do not assume a GUT
embedding, but instead require compatibility with a common high-scale
gauge coupling. Finally, the resulting \(\UB\) charges should remain
reasonably small while allowing the scalar and Yukawa interactions needed
for a viable phenomenological model.

For the non-supersymmetric models considered here, a complete mirror copy
of the SM matter representations degrades the gauge-coupling evolution.
The viable solutions therefore employ a reduced SM-like secluded sector.
A representative set of hypercharges satisfying these phenomenological
filters is shown in Table~\ref{tab:app_particle_content}.

\subsection{Explicit charge branches}
\label{app:explicit_charge_branches}

For the hypercharge assignments listed in
Table~\ref{tab:app_particle_content}, the anomaly equations can be solved
explicitly. A particularly simple branch is obtained by taking all signs in
Eq.~\eqref{eq:app_rationality} to be negative and choosing
\(\epsilon_Q=-\epsilon_L\equiv\epsilon\). For \(N=1\), the SM charges are
\begin{equation}
\begin{aligned}
    z_u &= -z_H-z_Q, \\
    z_d &=  z_H-z_Q, \\
    z_L &= -3z_Q-\frac{2}{3}q_S\epsilon, \\
    z_e &=  z_H+3z_Q+\frac{2}{3}q_S\epsilon, \\
    z_\nu &= -z_H+3z_Q+\frac{2}{3}q_S\epsilon .
\end{aligned}
\label{eq:app_sm_explicit_branch}
\end{equation}

The secluded-sector charges are then
\begin{equation}
\begin{aligned}
    q_L &= -3q_Q-2z_H+q_S\epsilon, &
    \tilde q_L &= 3q_Q+2z_H-2q_S\epsilon, \\
    q_{E} &= 3q_Q+z_H-q_S\epsilon, &
    \tilde q_{E} &= -3q_Q-z_H+2q_S\epsilon, \\
    q_{\nu} &= 3q_Q+3z_H-q_S\epsilon, &
    \tilde q_{\nu} &= -3q_Q-3z_H+2q_S\epsilon, \\
    q_{d1} &= -q_Q-z_H, &
    \tilde q_{d1} &= q_Q+z_H-q_S\epsilon, \\
    q_{d2} &= -q_Q+z_H, &
    \tilde q_{d2} &= q_Q-z_H-q_S\epsilon, \\
    q_Q &= q_Q, &
    \tilde q_Q &= -q_Q+q_S\epsilon.
\end{aligned}
\label{eq:app_explicit_charge_branch}
\end{equation}

The \(N=2\) branch has the same secluded-sector structure, while the
cancellation of the mixed \(SU(2)_L^2-\UB\) anomaly modifies the relation
between the SM lepton charge and the remaining parameters.

The baryonic benchmark studied in the main text is recovered by choosing the
parameters such that
\[
    z_Q=\frac13,\qquad
    z_u=z_d=-\frac13,\qquad
    z_L=z_e=z_\nu=z_H=0,
\]
for which \(\tB\) coincides with ordinary baryon number on the SM fields.


\begin{table}[t]
\centering
\small
\renewcommand{\arraystretch}{1.03}
\setlength{\tabcolsep}{3.5pt}
\begin{tabular}{lcccc}
\toprule
 & \multicolumn{2}{c}{Model A (\(N=1\))}
 & \multicolumn{2}{c}{Model B (\(N=2\))} \\
\cmidrule(lr){2-3}
\cmidrule(lr){4-5}
Field & \(U(1)_Y\) & \(\UB\) & \(U(1)_Y\) & \(\UB\) \\
\midrule
\multicolumn{5}{l}{\textit{Standard Model fields}} \\
\midrule
\(Q_L^f\)          & \(1/6\)  & \(1/3\)   & \(1/6\)  & \(2/3\)   \\
\(u_R^{c,f}\)      & \(-2/3\) & \(-1/3\)  & \(-2/3\) & \(-2/3\)  \\
\(d_R^{c,f}\)      & \(1/3\)  & \(-1/3\)  & \(1/3\)  & \(-2/3\)  \\
\(L_L^f\)          & \(-1/2\) & \(0\)     & \(-1/2\) & \(0\)     \\
\(e_R^{c,f}\)      & \(1\)    & \(0\)     & \(1\)    & \(0\)     \\
\(\nu_R^{c,f}\)    & \(0\)    & \(0\)     & \(0\)    & \(0\)     \\
\(H\)              & \(1/2\)  & \(0\)     & \(1/2\)  & \(0\)     \\
\midrule
\multicolumn{5}{l}{\textit{Secluded sector}} \\
\midrule
\(\psi_L^L\)        & \(-1/2\) & \(1/2\)   & \(-1/2\) & \(1/2\)   \\
\((\psi_R^L)^c\)    & \(1/2\)  & \(1\)     & \(1/2\)  & \(1\)     \\
\(\psi_L^E\)        & \(0\)    & \(-1/2\)  & \(0\)    & \(-1/2\)  \\
\((\psi_R^E)^c\)    & \(0\)    & \(-1\)    & \(0\)    & \(-1\)    \\
\(\psi_L^\nu\)      & \(0\)    & \(-1/2\)  & \(0\)    & \(-1/2\)  \\
\((\psi_R^\nu)^c\)  & \(0\)    & \(-1\)    & \(0\)    & \(-1\)    \\
\(\psi_L^{d1}\)     & \(1/3\)  & \(2/3\)   & \(1/3\)  & \(2/3\)   \\
\((\psi_R^{d1})^c\) & \(-1/3\) & \(5/6\)   & \(-1/3\) & \(5/6\)   \\
\(\psi_L^{d2}\)     & \(1/3\)  & \(2/3\)   & \(1/3\)  & \(2/3\)   \\
\((\psi_R^{d2})^c\) & \(-1/3\) & \(5/6\)   & \(-1/3\) & \(5/6\)   \\
\(\psi_L^Q\)        & \(1/6\)  & \(-2/3\)  & \(1/6\)  & \(-2/3\)  \\
\((\psi_R^Q)^c\)    & \(-1/6\) & \(-5/6\)  & \(-1/6\) & \(-5/6\)  \\
\(S\)               & \(0\)    & \(3/2\)   & \(0\)    & \(-3/2\)  \\
\bottomrule
\end{tabular}
\caption{\label{tab:app_benchmark_charges}
Two representative anomaly-free charge assignments. Model A is the benchmark
studied in the main text. Model B illustrates another solution in the same
class.}
\end{table}

Both assignments satisfy the anomaly equations and the high-scale
gauge-coupling criterion while keeping the largest charges reasonably small.

A complete phenomenological analysis of any given charge assignment requires
implementing the full set of renormalisable interactions, generating a
dedicated spectrum generator, and scanning the corresponding parameter space.
For this reason, the main text focuses on Model~A, while a systematic study of
the wider family of anomaly-free baryonic completions is left for future work.

\section{\SARAH model notation}
\label{app:dictionary}

Table~\ref{tab:sarah_dictionary} provides a correspondence between the notation used
throughout the paper and the parameter names adopted in the \SARAH
implementation.

\begin{table*}[t]
\centering
\small
\renewcommand{\arraystretch}{1.08}
\setlength{\tabcolsep}{4pt}

\begin{tabular*}{\textwidth}{
@{\extracolsep{\fill}}
l l
l l
@{}
}
\toprule
\multicolumn{1}{c}{Paper notation}
&
\multicolumn{1}{c}{\SARAH\ notation}
&
\multicolumn{1}{c}{Paper notation}
&
\multicolumn{1}{c}{\SARAH\ notation}
\\
\midrule

\(k_X\)
& \texttt{kX}
&
\(v_S\)
& \texttt{vS}
\\

\(k_{X1}\)
& \texttt{kX1}
&
\(v_X\)
& \texttt{vX}
\\

\(\lambda_{12}\)
& \texttt{lam12}
&
\(Y_d\)
& \texttt{Yd}
\\

\(\lambda_{112}\)
& \texttt{lam112}
&
\(Y_{d12}\)
& \texttt{Yd12}
\\

\(\lambda_A\)
& \texttt{lamA}
&
\(Y_e\)
& \texttt{Ye}
\\

\(\lambda_H\)
& \texttt{lamH}
&
\(Y_{e1}\)
& \texttt{Ye1}
\\

\(\lambda_{HS}\)
& \texttt{lamHS}
&
\(Y_{e12}\)
& \texttt{Ye12}
\\

\(\lambda_S\)
& \texttt{lamS}
&
\(Y_{l1}\)
& \texttt{Yl1}
\\

\(\lambda_{S1}\)
& \texttt{lamS1}
&
\(Y_\nu\)
& \texttt{Yv}
\\

\(\lambda_{SX11}\)
& \texttt{lSX11}
&
\(Y_{\nu1}\)
& \texttt{Yv1}
\\

\(\lambda_X\)
& \texttt{lamX}
&
\(Y_{\nu12}\)
& \texttt{Yv12}
\\

\(\lambda_{X1}\)
& \texttt{lamX1}
&
\(Y_{q12}\)
& \texttt{Yq12}
\\

\(\lambda_{X12}\)
& \texttt{lamX12}
&
\(Y_{sd}\)
& \texttt{Ysd}, \texttt{Ysu}
\\

\(\lambda_{X112}\)
& \texttt{lX112}
&
\(Y_{se}\)
& \texttt{Yse}
\\

\(\lambda_{XX1}\)
& \texttt{lamXX1}
&
\(Y_{sl}\)
& \texttt{Ysl}
\\

\(M_R\)
& \texttt{MR}
&
\(Y_{sq}\)
& \texttt{Ysq}
\\

\(\mu_H^2\)
& \texttt{muH2}
&
\(Y_{s\nu}\)
& \texttt{Ysv}
\\

\(\mu_S^2\)
& \texttt{muS2}
&
\(Y_u\)
& \texttt{Yu}
\\

\(\mu_{S1}^2=m_{S1}^2\)
& \(\texttt{muS12}=\texttt{mS1}^{2}\)
&
\(Y_{Xee}\)
& \texttt{YXee}
\\

\(\mu_{S12}^2=m_{S12}^2\)
& \(\texttt{muS122}=\texttt{mS12}^{2}\)
&
\(Y_{Xe\nu}\)
& \texttt{YXev}
\\

\(\mu_X^2\)
& \texttt{muX2}
&
\(Y_{X\nu\nu}\)
& \texttt{YXvv}
\\

\(v\)
& \texttt{vH}
&
\(y_{\rm off}\)
& \texttt{yoff}
\\

\bottomrule
\end{tabular*}

\caption{Dictionary between the notation used in the paper and in the
\SARAH\ implementation.}
\label{tab:sarah_dictionary}
\end{table*}


\bibliographystyle{utphys}
\bibliography{biblio}
\end{document}